Laser-assisted atom probe tomography of *c*-plane and *m*-plane InGaN test structures

N. A. Sanford[1], P. T. Blanchard[1*], M. D. Brubaker[1], A. K. Rishinaramangalam[2**], Qihua Zhang[3], A. Roshko[1], D. F. Feezell[2], B. D. B. Klein[3], and A. V. Davydov[4]

[1]National Institute of Standards and Technology, Physical Measurement Laboratory, 325 Broadway, Boulder, CO 80305

[2]Center for High Technology Materials, Electrical and Computer Engineering Department, University of New Mexico, Albuquerque, NM 87106

[3]Georgia Institute of Technology, Department of Electrical and Computer Engineering, Atlanta, GA 30332

[4]National Institute of Standards and Technology, Materials Measurement Laboratory, Gaithersburg, MD 20899

*Now with Honeywell, Broomfield, CO

**Now with Intel, Albuquerque, NM

## ABSTRACT

Laser-assisted atom probe tomography (APT) was used to measure the indium mole fraction *x* of *c*-plane, MOCVD-grown, GaN/$In_xGa_{1-x}N$/GaN test structures and the results were compared with Rutherford backscattering analysis (RBS). Four sample types were examined with (RBS determined) $x$ = 0.030, 0.034, 0.056, and 0.112. The respective $In_xGa_{1-x}N$ layer thicknesses were 330 nm, 327 nm, 360 nm, and 55 nm. APT data were collected at (fixed) laser pulse energy (PE) selected within the range of (2—1000) fJ. Sample temperatures were ≈ 54 K. PE within (2—50) fJ yielded *x* values that agreed with RBS (within uncertainty) and were comparatively insensitive to region-of-interest (ROI) geometry and orientation. By contrast, approximate stoichiometry was only found in the GaN portions of the samples provided PE was within (5—20) fJ and the analyses were confined to cylindrical ROIs (of diameters ≈20 nm) that were coaxial with the specimen tips. *m*-plane oriented tips were derived from *c*-axis grown, core-shell, GaN/$In_xGa_{1-x}N$ nanorod heterostructures. Compositional analysis along [0001] (transverse to the long axis of the tip), of these *m*-plane samples revealed a spatial asymmetry in charge-state ratio (CSR) and a corresponding asymmetry in the resultant tip shape along this direction; no asymmetry in CSR or tip shape was observed for analysis along $[\bar{1}2\bar{1}0]$. Simulations revealed that the electric field strength at the tip apex was dominated by the presence of a *p*-type inversion layer, which developed under typical tip-electrode bias conditions for the *n*-type doping levels considered. Finally, both *c*-plane and *m*-plane sample types showed depth-dependent variations in absolute ion counts that depended upon ROI placement.

1. Introduction

Laser-assisted atom probe tomography (APT) is gaining wide acceptance for 3-dimensional (3D), sub-nm-resolved, chemical mapping of metals, semiconductors, nanowires, superconductors, oxides, and biological materials.[1] In this paper we are only concerned with laser-assisted atom probe and not voltage-pulsed methods. Therefore, we adopt the abbreviation "APT" rather than "L-APT" –which is sometimes used to distinguish the laser-assisted approach from voltage pulsing. Very briefly, APT may be summarized as follows: a nano-needle shaped specimen is maintained at ≈ 50 K under ultra-high vacuum and biased at a voltage that is held just below the threshold for field evaporation of ions. The corresponding electric field strength (on the vacuum side of the vacuum/tip interface) to establish this condition is on the order of 10 V/nm. Field evaporation is then triggered by a pulsed laser incident upon the specimen. The mechanism is generally assumed to arise from a thermally-driven pathway whereby transient temperature increases induced by the laser momentarily reduce the threshold for field evaporation at the specimen apex and ions are then emitted synchronously with the laser pulse frequency. One should not confuse APT with laser ablation; APT generally employs incident laser pulse energies that are many orders of magnitude below the threshold for ablation. Ions field-evaporated from the specimen and accelerated in the applied electric field are picked up by a two-dimensional detector which enables recording their impact locations and times-of-flight. In principle, the process may proceed indefinitely but practical considerations often limit data sets to the collection of between roughly $10^5$ to $10^8$ ions and analysis depths to ≈ 1 μm. The accumulated data is used to calculate a 3D chemical map, or "reconstruction," of the specimen by computing the original location of each elemental specie with the corresponding ion identified by its time-of-flight. Thorough and comprehensive reviews of APT theory and methods have been published.[2,3,4]

Closer to the topic at hand, Rigutti, et al (Ref. 5) have surveyed APT analysis of wide-bandgap, III-N compound semiconductor device structures, which includes LEDs, lasers, and high electron-mobility transistors. As they succinctly noted, "*Due to the relatively high rate of success for the analysis and to the straightforward interpretation of their mass spectra, III-N materials have become a model system for the study of compositional biases in APT, which may occur in a much broader class of compounds,*" (emphasis added). Studies of APT compositional biases in other wide-bandgap materials, including ZnO and MgO, have also been reported.[6,] We interpret and define the phrase, "compositional biases," as the departure of APT-derived, spatially-resolved chemical mapping of a specimen from the specimen's true, physical nature and composition. Common shortcomings of APT that arise in studies of III-N materials, and related compositional bias effects, are discussed next.

Reconstructions (abbreviated herein as, “recons”) derived from APT can depend upon the operational conditions of the instrument. To illustrate perhaps the simplest case for GaN: specimen longevity is promoted by using a relatively high laser pulse energy (PE). Elevated PE permits operation at a relatively reduced specimen bias voltage, which reduces the likelihood of specimen fracture while maintaining a desired detection rate (DR) of ions. However, elevated PE applied to GaN will often return a compositional bias that indicates an (unphysical) deficiency of nitrogen.[5,6,7] The reverse can also occur ─ a relatively low PE can return a composition bias rich in N. Such compositional biases are unphysical in GaN because the material exists as a line compound and will thus decompose if its true stoichiometry deviates even slightly from 50% Ga and 50% N.[8] Consequently, situations involving, say, APT analysis of a particularly fragile GaN heterostructure specimen may *force* an operator to choose data acquisition conditions in a regime where it is known beforehand that the returned concentrations of Ga and N will be incorrect—which may immediately call into question the analytical accuracy of APT for determining the concentration and distribution of the various other alloy constituents and dopants.

APT applied to the study of III-nitrides and other wide-bandgap materials has evolved from employing a laser operating in the visible (laser wavelength λ = 532 nm, Ref. 9) and now more typically uses near-ultraviolet lasers, e.g., λ = 343 nm or 355 nm.[6,7] Evidence also suggests that operating at even shorter wavelengths, notably λ = 258 nm, should yield superior results.[10] Nonetheless, compositional biases of varying degrees are obtained with all of these conventional lasers. Controversy also exists as to the nature of the underlying pathways, thermal or otherwise, responsible for laser-assisted field evaporation. For example, an APT study (λ = 355 nm) of MgO ascribed the photogeneration of holes and surface states as primary mechanisms for field evaporation since λ was well below the bulk bandgap (≈ 7.7 eV) for MgO.[11] For GaN with λ=355 nm, Diercks, et al. (Ref. 12) argues for the possibility an athermal field evaporation mechanism—which is interesting considering that 355 nm corresponds closely to the low-temperature absorption edge of bulk GaN (≈ 3.48 eV).[13,14] More recently, a photoionization pathway for APT has been proposed; this approach would replace the conventional, near-UV laser with an extreme-UV (EUV) laser.[15,16] A prototype EUV-equipped APT tool with λ = 29.6 nm (42 eV) was indeed constructed and initial results on $SiO_2$ (amorphous fused silica) showed recovery of the correct stoichiometry.[17] Finally, besides correlation with the particular laser wavelength and PE, compositional biases may also be influenced by the presence of defects in the specimen, and the crystallographic direction along which the analysis was performed.[5,18,19]

Clearly, there are numerous unresolved issues associated with the mechanisms responsible for both field evaporation and compositional biases in APT. However, perhaps the main question confronting a materials engineer supporting III-nitride development boils down to: "If I operate my atom probe tool in a fashion that insures analysis to the required depth without fracturing the specimen, what can I expect in terms of spatially-resolved analytical sensitivity, precision, and uncertainty for measurements of the dopants and alloy constituents of interest?" This paper is primarily addressed to workers posing that question.

In our study of GaN/$In_xGa_{1-x}N$/GaN multilayer samples we have compiled APT data that was collected under various operational conditions. For *c*-plane, $(0001)$, specimens taken from samples of planar epitaxial films, the values of *x* derived from APT are compared to results obtained from Rutherford backscattering analysis (RBS) performed on the same samples. For *m*-plane, $(10\bar{1}0)$, quantum well (QW) specimens taken from core-shell micropost samples, the APT-derived results for *x* are compared with values of *x* that have been previously calibrated by photoluminescence studies at the University of New Mexico. Variously, for both *c*-plane and *m*-plane samples, the (APT) reconstructed layer thicknesses are calibrated by means of separate X-ray diffraction (XRD) and transmission electron microscopy (TEM) measurements. A key contribution of this paper for the *c*-plane samples is to obtain APT tool operational conditions that return indium concentrations that conform, within experimental error, to RBS. We then apply the RBS-anchored tool conditions to APT analysis of the *m*-plane samples and find that the indium concentration in the QWs conforms to what is expected from photoluminescence.

Interestingly, we observed an asymmetric, spatially-varying artifact in GaN charge state ratios (CSRs) for the *m*-plane GaN specimens when the analysis direction is along [0001], i.e., transverse to the long axis of the specimen tip. The effect is absent when the analysis is performed in the non-polar $[\bar{1}2\bar{1}0]$ direction; it is also absent when analysis is performed in non-polar directions on *c*-plane tips. As will be discussed at length in Section 3.1.2, the feature is attributed to an observed asymmetric evolution of the apex of an *m*-plane tip such that the apex displaces toward the -*c* direction as ions are field evaporated away. By contrast, *c*-plane tips generally evolve symmetrically during APT.

In our terminology, "specimen(s)" or "tip(s)", and "lamella(e)" are prepared from "samples" for examination by APT and TEM, respectively. A focused ion beam (FIB) tool, which incorporates a field-emission scanning electron microscope (FESEM), was used to prepare all tips and lamellae. Finally, we make interchangeable use of the terms "measured composition and "composition" in referring to APT results—and emphasize that such output will often differ from the true, *physical composition* of the specimen at hand.

Following this introduction, the paper is organized as follows: Section 2 describes the samples used in this study and specimen preparation therefrom. Section 3 presents the results of APT performed on a series of specimens taken from *c*-plane GaN/$In_xGa_{1-x}N$/GaN multilayer samples. Analysis of specimens prepared from an *m*-plane, quantum well (QW) sample is also presented. A brief summary of the electrostatic analysis is presented in Section 4 with a more extensive development of such simulation work deferred to a separate, upcoming publication. A discussion of experimental and analytical uncertainties is included in Section 5 and conclusions are presented in Section 6.

## 2. Samples examined

### 2.1 *c*-plane GaN/$In_xGa_{1-x}N$/GaN multilayers

This collection of samples consists of 4 separate MOCVD growth runs with respective RBS-determined values of *x* of: 0.112, 0.056, 0.034, and 0.030; the corresponding layer thicknesses, as variously determined by XRD or growth rate, are given in Table 1. These samples were used in an entirely separate study discussed in Ref. 20 and preliminary RBS results were given therein; the RBS analysis presented in this paper was separately performed, but the results are quite similar.[21] APT specimen tips were prepared using FIB techniques that have been discussed at length elsewhere.[22,23] FIB techniques for preparing TEM lamellae are well established and generally documented in FIB operation manuals.

### 2.2 *m*-plane GaN/$In_xGa_{1-x}N$/GaN QW structures

These samples were arrays of *c*-axis oriented, MOCVD-grown, core-shell microposts with QWs grown on the *m*-plane sidewall facets; fabrication details are described elsewhere.[24,25,26] The micropost core was comprised of lightly-doped (free carrier concentration in the mid-low $10^{17}cm^{-3}$), *n*-type GaN grown using pulsed-MOCVD to achieve a high vertical growth rate and minimal lateral growth. The growth was performed under a $H_2/N_2$ mixed atmosphere at 13.3 kPa and 940°C. The V/III ratio employed during the growth was ~100. The growth of the core section was followed by the growth of three pairs of InGaN/ GaN quantum-well shells around the posts in an $N_2$ atmosphere using a high V/III ratio (~10,000). Figure 1 (a) illustrates an array of these structures and Fig. 1 (b) shows a TEM image that illustrates the thicknesses of GaN capping layer, $In_xGa_{1-x}N$ QWs, and GaN barrier layers. Both the APT specimens and TEM lamellae were fabricated using FIB techniques that were very similar to those used for the *c*-plane samples.

## 3. APT analysis of samples

### 3.1. Overview

APT data were acquired with a LEAP 4000XSi atom probe tool manufactured by CAMECA Instruments; data analysis was performed using the IVAS software package provided by the same company.[27] For all cases considered in this paper, the specimen-detector flight length was 90 mm. The laser installed on the tool operates with a pulse width of ≈ 10 ps, at a wavelength $\lambda$ = 355 nm, and is focused to a diameter of ≈ 2 μm at the specimen tip. Tool operational parameters are abbreviated as: laser pulse energy (PE), laser repetition rate ( $f$ ), data detection rate (DR), which is ions/pulse (expressed as a percentage), and sample temperature (T). In our usage, "voltage," "specimen voltage," and "bias voltage" refer to the electrode-tip voltage bias that produces field evaporation of ions from the specimen. In this context, "field evaporation," is defined by Miller.[28] Additionally, IVAS data analysis outputs and parameters are abbreviated as: region of interest (ROI), top-level ROI (TLROI), detector efficiency (DE), image compression factor (ICF), and sphere-cone-radius-ratio (S/C). For all recon cases presented, we used the "Tip Profile" method in the IVAS software whereby an FESEM image of the as-FIB-prepared specimen is imported into the software. Additionally, an FESEM image of the tip shape after data acquisition was used to estimate S/C for inclusion in the IVAS analysis. For all data acquisition cases presented in this paper, $f$ = 250 kHz.

Mass-spectral peak assignments are given in Table 2. As will be discussed in the following sections, the relative (and absolute) detected quantities of these species depend upon tool operational conditions and selection of ROI. We adopt the simplifying convention that the peak at 14 Da is assigned to the $N_2^{2+}$ ion.[7] Universally assigning 14 Da only to $N_2^{2+}$ is not strictly valid since an admixture, or predominance, of $N^{1+}$ may certainly exist, but there is not enough information to unequivocally deconvolve and weigh the separate contributions. Nonetheless, we find the overall uncertainty penalty emerging from such a restrictive and presumptive assignment is not particularly severe. Additionally, we assume that peaks in the (15 —18) Da range arise from complexes of $N_2$ and H. Moreover, counts arising from H, $H_2$, $H_3$ assignments, and H derived from the (IVAS) decomposition of $N_2$ —H complexes, are all assumed as background, contribute no volume to a reconstruction, and are not included in compositional analyses. Finally, over the range of $x$ considered in this paper, the volume of the $In_xGa_{1-x}N$ primitive cell does not substantially deviate from that of GaN—with regards to its influence on computing reconstructions. In particular, the volume contributed per ion for GaN is 0.0114 nm$^3$ as deduced from tabulated XRD data.[29] The highest RBS-determined indium concentration we consider in this paper is $x \approx 0.11$ (sample h). Our XRD analysis, in conjunction with the results of Ref. 30, allow us to estimate the volume occupied per ion is 0.0117 nm$^3$ for the case of $x \approx 0.11$. Therefore, for simplicity, we assign the volumetric contribution per ion to 0.0114 $nm^3$ for all samples considered since correcting for the crystallographic cell volume of even our highest indium-containing sample will have a negligible effect on the reconstructions. Consequences of the foregoing assumptions on estimated uncertainties of the APT results are discussed further in Section 5.

3.1.1. *c*-plane, GaN/$In_xGa_{1-x}N$/GaN multilayer samples

We begin by discussing sample h in some detail since the (≈ 55 nm) thickness of the associated InGaN layer provides a marker for constraining the selection of recon parameters. Such constraints then offer justification for selecting parameters used for the analyses of the other multilayer samples. For samples f, d, and c, the InGaN layer thicknesses are inconveniently large and excessive APT operational time is required to penetrate the entire InGaN layer thickness. In such cases of relatively thick layers, both top and bottom InGaN/GaN interfaces cannot be used as markers with which to constrain the choice of recon parameters. Tables 3—6 give a detailed enumeration of all the recon cases for samples h, f, d, and c, respectively. In the discussion that follows for the *c*-plane samples, the (*x*, *y*, *z*) coordinate system of a recon follows the IVAS convention and the *z* axis is nominally collinear to the crystallographic *c* axis of a specimen. However, in the IVAS convention, +*z* direction indicates depth (from tip apex to tip base) into the specimen. Our *c*-plane samples are grown with Ga-polarity. The crystallographic +*c* direction, i.e. [0001], for a resulting specimen tip is therefore antiparallel to the IVAS *z*-axis.

Figures 2(a—c) illustrates the evolution of the (sample h) specimen tip of recon 1, which shows: (a) the FIB-prepared tip that supports a thin Ni capping layer, (b) the resulting tip after the Ni cap and a portion of the GaN top layer have been field-evaporated away, and (c) the remaining tip after the InGaN layer, and a portion of the underlying GaN layer, have been field-evaporated away. Estimation of S/C for this tip is thus obtained from Fig. 2(c). The range and variation in tip voltage corresponding to the data used recon 1 is given in Fig. 3, and the aggregate mass spectrum for all the ions accumulated over the same voltage range is shown in Fig. 4.

Further details of recon 1 are now presented. Figure 5(a) shows the 3D chemical map illustrating the InGaN layer embedded between top and bottom GaN regions. In IVAS nomenclature, this map is the “top-level ROI” (TLROI), which incorporates all ions collected over the voltage span utilized. With parameters indicated in Table 3, IVAS simultaneously renders both InGaN/GaN interfaces as approximately planar and parallel; the IVAS-derived InGaN layer thickness is within ≈ 1 nm of the result obtained by XRD. The InGaN layer appears tilted with respect to the z axis because FIB processing variations prevented a more nearly perpendicular orientation of tip and sample. A graph of the 1D, TLROI, *z*-axis concentration profile for Ga, N, and In, corresponding to Fig. 5(a), is shown in Fig. 5(b); the Ga/In transitions appear artificially graded, which results from the tilted orientation of tip as just described. Figure 5(b) also illustrates a non-physical, N-rich composition of both the top and bottom GaN layers. Figure 6(a) shows a 20-nm-diameter, cylindrical ROI that subdivides the TLROI; a z-axis concentration profile constrained to this ROI is shown in Fig. 6(b). The results illustrated in Figs. 5(a, b) and 6(a, b), and listed in Table 3 for this PE = 2 fJ case, indicate that the APT derived concentrations for In, N, and Ga within the InGaN layer are quite insensitive to ROI diameter. On the other hand, the measured concentrations for Ga and N in the GaN regions are comparatively dependent upon the geometry of the ROI. We now explore concentration profiles along ROI axes oriented perpendicular to the TLROI *z*-axis of recon 1.

Figure 7 shows a 2D relative-density map for Ga that exhibits 6-fold rotational symmetry (about z). This map is derived from recon 1 and plotted in the coordinate system of the reconstruction such that the relative density of Ga is projected onto the *x*-*y* plane at the maximum *z*-depth of the TLROI. Maps analogous to Fig. 7 were noted as revealing the 6-fold symmetry of the $\langle 0001 \rangle$ zone of the III-N wurtzite lattice.[7] We use Fig. 7 to place two cuboid-shaped ROIs (ROI A and ROI B), both oriented perpendicular to *z*, and both intersecting the TLROI as illustrated in Fig. 8. ROIs A and B are placed to intersect regions of high and low relative Ga density, respectively. The dimensions of these ROIs are indicated in the associated figure captions. Figures 9 (a, b) show that the comparative axial concentration profiles (for N, Ga, and In) along A and B are nearly equal—regardless of what Fig. 7 may suggest to the contrary. However, as described next, these effects are complex and the spatial distribution of measured composition can depend upon PE, the ionic charge state represented, and may not reveal any underlying crystal symmetry whatsoever.

The data of Fig. 9 (a) is recast in Fig. 10 (a, b) to illustrate the 1D concentration profiles for the prominent charge states observed for Ga and In. Recasting Fig. 9 (b) yields similar results, which are omitted for brevity. Figure 10 (a) shows the concentration profile of $Ga^{1+}$, and the combined profiles of $Ga^{2+}$ with $Ga^{3+}$ (abbreviated as $Ga^{2,3+}$, similarly for $In^{2,3+}$). Notably, $Ga^{1+}$ falls nearly to zero in the center. $Ga^{1+}$ and $Ga^{2,3+}$ vary with similar magnitudes, but in opposition, such they add up to the relatively uniform Ga concentration profile shown in the graph of Fig. 9 (a). A similar complementary trend for $In^{1+}$ and $In^{2,3+}$ is shown in Fig. 10 (b), but in this case (PE = 2 fJ) the $In^{1+}$ contribution is comparatively minor.

We now survey the results of recon 3 for the case of PE = 10 fJ. Figure 11 (a, b) show graphs of the respective 1D, *z*-axis concentration profiles for the TLROI and a 20-nm-diameter, coaxial ROI cylinder. For brevity, the corresponding 3D representation of the TLROI is omitted. Both graphs indicate nonphysical composition trends for N and Ga but the effect is more pronounced for the GaN region of Fig. 11 (a). However, Fig.

11 (b) illustrates that roughly stoichiometric Ga and N fractions are found in the GaN region when the analysis is constrained to the coaxial ROI cylinder.

In a similar manner as just discussed, Figs. 12 (a, b) compare concentration profiles between the TLROI and a 20-nm-diameter, coaxial ROI cylinder for the case of recon 5 with PE = 100 fJ. The deviation from stoichiometry in the GaN region is quite pronounced in both cases. However, the 4 examples illustrated in Figs. 11 (a, b) and 12 (a, b) indicate that the indium concentration is comparatively insensitive both the ROI selection and PE.

Figures 13 (a, b) illustrate transverse concentration profiles for recon 5 for the case of PE = 100 fJ. The ROI is denoted by "A" and has the same dimensions, and approximate orientation (with respect to the laser input) as for the transverse ROI (pertaining to recon 1) shown in Fig. 8. For this case, however, the corresponding 2D map of relative Ga density shows an approximately radially symmetry; the 6-fold azimuthal feature, as seen in Fig. 7, is not evident for recon 5. Hence, for Figs. 13 (a, b), "radial concentration fluctuations" and "transverse concentration fluctuations" are synonymous. Figure 13 (a) illustrates that both $Ga^{1+}$ and $Ga^{2+}$ are detected, but $Ga^{1+}$ is the dominant specie. The net Ga concentration is not radially uniform but is lower at the center of the TLROI than at the sidewalls. The radial variations of the $In^{1+}$ and $In^{2+}$ concentrations appear in Fig. 13 (b); the $In^{3+}$ charge state was not detected. Interestingly, and in contrast to the case of recon 1, Fig. 13 (b) illustrates that the radial concentration fluctuations of the indium species are of similar magnitudes, but in opposition, such that the net transverse indium concentration adds up to be roughly uniform. Similar results are found in Fig. 14 (a, b) that pertain to ROI "B".

The results surveyed in Figs. 5—14 illustrate that the 3D distribution of chemical composition, and the charge states of individual species, can depend strongly on PE, and necessarily, voltage. For example, on the one hand Figs. 13 and 14 illustrate an overall spatially-dominant contribution of $Ga^{1+}$, which might be expected to occur (at this relatively increased 100 fJ PE case) from post-ionization theory.[31] On the other hand, the figures also show that the measured $In^{1+}$ and $In^{2+}$ concentrations vary spatially and are about equal in the center of the ROIs. Evidently, the adherence of these results to post-ionization theory for this ternary compound is qualitative at best. Given such behavior, and the need to minimize the probability of specimen fracture during data acquisition, the importance of tabulating APT results that utilize data collected under various SV and PE conditions, and are analyzed within different ROIs, is immediately evident. In Section 5.4, we discuss results for the computed electric field strength at a tip's apex and compare our measured CSR data for Ga with results of post-ionization theory applied to GaN.[6]

Table 3 summarizes the results for sample h. Here, ROI 1 encompasses the entire InGaN layer but omits the top and bottom GaN regions. ROI 2 is a 20-nm-diameter cylinder, coaxial with the TLROI, and extending over the InGaN layer thickness. The column headings for N, Ga, and In refer to the integrated concentrations (at. %) within the ROIs for the respective species. In all cases, the IVAS-computed concentration, with background subtracted, of a specie included both elemental assignments and contributions that were software-decomposed from complex molecular ions. For specimen temperatures of 54 K and 26 K, results are shown for several PE cases ranging over (2—1000) fJ and (2—50) fJ, respectively. Additionally, the IVAS fitting parameters used for each recon are also given in the table. Interestingly, for a given PE and temperature, the indium concentration is seen to be rather insensitive to the choice of ROI. The key trends that emerges from the results are: (i) The measured concentration of indium increases with increasing PE. (ii) There is no significant difference between the indium concentrations computed for the two temperature cases. (iii) PE in the approximate range of (10—50) fJ will produce indium concentrations that are in closest agreement with RBS. We now present results for the remaining *c*-plane, GaN/$In_xGa_{1-x}N$/GaN, multilayer samples.

Tables 4, 5 and 6 summarize the L-APT analyses of samples f, d, and c, respectively. For all three samples we again observe that neither the choice of ROI nor specimen temperature make a significant difference on measured indium concentration, and that the indium concentration increases with increasing PE. For these respective samples, however, the corresponding approximate PE ranges that most closely conform to the RBS results are, (2—5) fJ, (2—10) fJ, and (2—50) fJ. In examining samples f, d, and c, each supporting thick InGaN regions compared to sample h, it was impractical to run the APT data acquisition long enough to penetrate the full thickness of the InGaN layers. Therefore, only the initial and final tip shapes, the residual tip length, and the upper GaN/InGaN interfaces, were available to guide the selection of recon parameters.

3.1.2 *m*-plane GaN/$In_xGa_{1-x}N$/GaN QW samples

These QW specimens were comparatively fragile and fractured easily during APT. Thus, it was not practical to accumulate a wide range of PE cases that would simultaneously render all 3 QWs as illustrated if Fig. 1(b). Additionally, the GaN capping layer was ≈ 5 nm thick and provided little material for the preliminary tip-formation during the alignment phase of APT data acquisition. Therefore, the top-most QW was often consumed during the alignment process, or had been partially milled away by the FIB, and only the bottom two QWs usually provided useful data. Furthermore, the nature of these specimens, as derived from arrays of *c*-axis grown GaN/InGaN core-shell microposts, precluded RBS analysis. Nevertheless, it is instructive to compare L-APT performed on these *m*-plane, core-shell QW samples to the results obtained from thicker, *c*-plane, “bulk like” InGaN layers. It is also instructive to compare these core-shell specimens with prior APT work on planar, *m*-plane GaN material as reported by other workers.[18]

Table 7 lists details of APT analyses of three representative cases for the core-shell QW specimens. PE= 10 fJ in recon 30 and all 3 QWs were recovered. In recons 31 and 32 the respective PE values were 15 fJ and 20 fJ and only the lower 2 QWs were recovered for those cases. The listed IVAS fitting parameters were selected as compromises to approximately duplicate the TEM-determined separation between the 1$^{st}$ and 3$^{rd}$ QWs for recon 30, the separation between the 2$^{nd}$ and 3$^{rd}$ QWs for recons 31 and 32, and render QW layers as approximately planar in each case. As discussed next, simultaneously meeting these constraints in the reconstruction analysis, even for QW layers that are comparatively thin and closely spaced as these were, is usually not achievable.

Figure 15 (a) shows the TLROI for recon 30 and reveals two recurring problems: (i) FIB processing issues have again yielded a tip that is tilted with respect to the QW layers, (ii) the IVAS software does not simultaneously render the 3 QWs as approximately planar. Similar issues were encountered for other reconstructions (not shown) of this sample, i.e., a tilted tip and not-simultaneously-planar QWs. Figure 15 (a) also shows a 15-nm-diameter ROI cylinder, placed roughly perpendicular to the QWs, and Fig. 15 (b) shows the 1-D concentration profile along the axis of this ROI. Taken together, these results again show that the IVAS fitting parameters, reconstructions, and the placement and diameters of the cylindrical ROIs, represent compromises in efforts to duplicate the TEM-determined QW separations and render approximately planar layers. Comparatively speaking, considering all samples examined in this study, it was quite unusual that (roughly) planar and parallel interfaces *could* be rendered for sample h as illustrated in Fig. 5(a).

Figure 15(c) compares the *z*-axial, In concentration profiles generated within 15-nm-diameter ROI cylinders for the 2$^{nd}$ and 3$^{rd}$ QWs for the 3 PE cases of recons 30, 31, and 32. In each case, the ROI cylinders are oriented roughly perpendicular to the QW layers. The results of this (small) set of measurements suggest that there may a resolvable trend whereby increasing PE from 10 fJ to 20 fJ results in a corresponding fractional increase in measured indium concentration of roughly 10%. However, this cannot be unambiguously distinguished from possible spatial variations of In concentration that may occur between separate nanorods during growth. Photoluminescence was also used to separately estimate the QW alloy composition in these micropost samples and yielded $x \approx 0.10$, which, for this PE range, is consistent with APT studies of the *c*-plane samples.

Figure 16 illustrates a 2D relative-density plot for Ga that was generated under recon 32 with PE = 20 fJ. In contrast to Fig. 7, however, Fig. 16 shows no azimuthally 6-fold rotational symmetry, but instead shows an in-plane, 2-fold axis of symmetry. Notably, the 2-fold feature is not "washed out" at the PE used. As discussed above, the 6-fold feature shown in Fig. 7 with PE = 2 fJ did not persist in recon 5 where the PE was 100 fJ, which illustrates that the appearance of these underlying symmetry features very much depends on the PE.

Recon 31 is useful to illustrate this 2-fold effect in relative Ga density since the tip had the least-tilted orientation, due to FIB mounting errors, with respect to the QW/GaN interfaces compared to all other core-shell specimens examined. Given our specimen mounting convention, and the separately-evaluated polarity of the core-shell nanoposts [Refs. 24—26], the 2-fold axis shown in Fig. 16 conforms to the crystallographic *c*-axis of the sample with the [0001] direction as indicated. The correlation of the *c*-axis polarity with the character of the 2D map of Fig. 16 is consistent with that reported in Ref. 18. If we assume that the data acquisition and reconstruction algorithms faithfully retain the physical orientation and symmetry of the specimen, then the reader views the 2D map of Fig. 16 by looking along the $[10\bar{1}0]$ direction; the $[\bar{1}2\bar{1}0]$ direction is thus established as shown. The common Miller-Bravais indexing convention for a hexagonal lattice geometry is given in Ref. 32.

Figure 17 (a) illustrates the TLROI corresponding to recon 31. The figure also shows two ROI cylinders placed along [0001] and $[\bar{1}2\bar{1}0]$, respectively. Note that both ROI cylinders are 15 nm in diameter and intersect the TLROI only in the GaN region at a depth of roughly 50 nm below the deepest QW. Figure 17 (b) shows the axial concentration profile of Ga and N within the ROI along $[\bar{1}2\bar{1}0]$; Fig. 17 (c) show the concentration profile for $Ga^{1+}$, $Ga^{2+}$, and N along the same axis. Both graphs are essentially radially symmetric with respect to the TLROI. By contrast, as shown in Figs. 17 (d) and (e), the concentration profiles for Ga, N, $Ga^{1+}$, and $Ga^{2+}$ for the ROI placed along [0001] are noticeably asymmetric. Figure 17 (f) shows the final state of the tip after the completion of the APT run and illustrates that the apex is displaced in the -*c* direction and away from the axial centerline of the tip. We now correlate these directionally dependent trends in CSR with both the symmetry of the tip and the strength of the surface electric field.

By principles of basic electrostatics, SV bias of a uniform GaN tip will produce a maximum surface field $E_a$ at the tip apex.[33,34] In our usage, "surface field" or "surface electric field" synonymously refer to the electric field on the vacuum side of the tip-vacuum interface. Therefore, the transverse concentration profiles for $Ga^{1+}$ and $Ga^{2+}$ can be regarded as transverse spatial indicators of the electric field strength across the surface of the tip with a maximum in $Ga^{2+}$ counts occurring at the geometrical apex. This trend is indeed revealed in both *c*-plane and *m*-plane specimen tips but is most striking in *m*-plane specimens where it is radially asymmetric when viewed along [0001].

Specifically, when an *m*-plane tip is viewed in FESEM after APT analysis such that the [0001] axis is in the plane of the image, the resulting, quasi-hemispherical apex region is asymmetric, and the apex itself is displaced in the $[000\bar{1}]$ direction. Correspondingly, the relative concentration of $Ga^{2+}$ is also spatially asymmetric and displaced in the $[000\bar{1}]$ direction. On the other hand, if a tip is viewed with a non-polar axis in the plane of the image, both the tip and the relative concentration profile of $Ga^{2+}$ along that axis appear spatially symmetric. The tip evolution shown in Fig. 2, for a *c*-plane tip, combined with the associated graphs of Ga CSR of Figs. 10 (a, c), are representative of symmetric tip evolution.

However, an obvious caveat to keep in mind is that a tip will evolve asymmetrically for trivial cases when FIB processing inadvertently results in a slightly tilted orientation. Using the IVAS software, such mounting errors can be immediately recognized as simply *x-y* displacements from symmetry of the 2-fold or 6-fold relative intensity maps for the respective *m*-plane and *c*-plane specimens. This effect can be seen in Fig. 7 for a slightly tilted FIB mounting of the *c*-axis tip since the center of the 6-fold feature is displaced in the positive *x-y* direction. On the other hand, the *m*-axis tip of Fig. 16 appears to be mounted without any significant tilt since the 2-fold map is symmetrically disposed. Hence, the correlation between apex evolution and *c*-axis orientation in *m*-plane specimens may be obscured and overlooked due to accidental FIB mounting artifacts.

A possible hypothesis to explain the asymmetric evolution of the *m*-plane tips is to associate it with the spontaneous polarization field $\vec{P}_S = -\hat{c}|\vec{P}_S|$ C/m$^2$, which is known to exist in GaN. Values of $|\vec{P}_S|$ ranging from 0.007 C/m$^2$ to 0.034 C/m$^2$ have been reported.[35,36,37] In this convention, $\hat{c}$ is a unit vector directed along [0001] and $\vec{P}_S$ points from the Ga face toward the N face of the crystal. The magnitude of the electric field produced by this dipole is therefore (very roughly) $|\vec{P}_S|/\epsilon_o$ or ≈ (0.8—3.5) V/nm.[33,34] Then, since we may estimate $E_a$ to fall in range of (10—20) V/nm [Ref. 3], the dipole field would be a significant transverse perturbation to $E_a$. Therefore, the vector sum of these fields could compel the tip to evolve with an asymmetrically displaced apex as ions are field evaporated away. Although intuitively attractive, this viewpoint is dismissed in Section 4 where we present that under typical bias conditions, the apex of the semiconducting GaN tip is generally in a state of inversion and the resulting density of mobile, free holes would screen any effect $\vec{P}_S$ would have on $E_a$. Instead, as discussed next, we propose that the asymmetric tip evolution is expected from growth habits, etch behavior, and Wulff plot analysis of *m*-plane GaN.[38,39,40]

Jindal, et al [Ref. 38] performed selective area growth of GaN on $(1\bar{1}00)$ GaN substrates and produced nanostructures with "arrow-headed shapes" such that the tip and base of an arrowhead point toward the [0001] and $[000\bar{1}]$ directions, respectively. The growth velocities in the $[000\bar{1}]$ and directions $[1\bar{1}00]$ are lower than that of the [0001] and $[1\bar{1}01]$ directions, producing a structure that is anisotropic in the *c*-direction. Consequently, as illustrated in Fig. 18, the geometric apex of such a nanostructure is displaced toward in the $[000\bar{1}]$ direction but the apex is symmetric with respect to the *a*-axis. Additionally, Megalini [Ref. 39] has performed photo-electrochemical (PEC) etching of *m*-plane GaN structures and observed that the (0001) Ga face etches faster than the $(000\bar{1})$ N face. Taken together, these finding suggest that it is not unreasonable to expect that our *m*-plane APT tips should evolve in the asymmetric manner described previously and illustrated in Fig. 17 (f). Moreover, Fig. 18 also qualitatively suggests that when viewed with the *a*-axis in the plane of the image, the evolution of an *m*-plane tip should be symmetric, but also wider, than for the *c*-axis in the plane of the image. Figure 19 indeed illustrates this trend for the case of the *m*-plane GaN tip of recon 33.

## 4. Electrostatic analysis of GaN specimen tips

Our electrostatic analysis will consider only the representative case for the *m*-plane tip of recon 31 but confined to the GaN portion after the QWs have been field evaporated away. This abbreviated presentation is instructive and also puts into context application of the common "*k*-factor" approximation [Ref. 3] for computing $E_a$ when semiconductor specimens are considered. A more comprehensive electrostatic study is under development for presentation elsewhere.[41]

A schematic of the tip-electrode geometry is given in Fig. 20. The key assumptions are as follows: The Si coupon and micropost are degenerately doped and treated as metallic conductors. The micropost stands 200 μm above the coupon and has the same taper as the FIB-milled specimen tip. As found from recon 31, the residual length of the GaN tip protruding from the Si micropost is ≈ 480 nm; this is the remainder of the tip after the quantum well layers have been removed. The tip apex diameter, as measured by FESEM following APT data acquisition, is ≈ 81 nm. For simplicity, the GaN tip is assumed to be ohmically bonded to the Si post. The nominal LE dimensions chosen and shown in Fig. 20 are adapted from other publications.[3,42]. The electrode-tip separation of ≈ 40 μm was experimentally estimated.

### 4.1 Isotropic approximation

The inherent spontaneous polarization of the GaN specimen is ignored. We assume the material has a dielectric constant of 8.9 and is isotropic. The realistic estimate for the *n*-type free carrier concentration used is $1\times10^{17}$cm$^{-3}$. The tip-electrode voltage is set to 4 kV in conformance to the approximate voltage at the end of data acquisition sessions for recon 31. A finite-element Poisson solution scheme was developed and implemented. The results show that the *n*-type specimen tip is depleted of bulk free carriers for SV ≈ 4 kV. However, the 4 kV is strong enough to induce an inversion layer of free holes in a region of the tip encompassing the apex. The conducting inversion layer would effectively screen the applied electric field from the interior of the specimen and the resulting on-axis, maximum surface electric field $E_a$, (again, on the vacuum side of the vacuum/apex interface) is ~ 24 V/nm. Moreover, the conducting inversion layer would also screen a tangential dipole field on the apex, arising from the built-in spontaneous polarization. Therefore, as described previously, a transverse displacement or perturbation of $E_a$ due to $\vec{P}_S$ may be eliminated from consideration in our electrostatic analysis.

In regarding the biased GaN tip as a metallic conductor, rather than a semiconductor, and following Ref. [3], $E_a$ may be estimated by what we refer to as the "*k*-factor approximation". Within this approximation, $E_a = V_o / k\rho$ where $V_o$ is the electrode-tip voltage, $\rho$ is the radius of curvature of the tip, and $k$ is a dimensionless fitting parameter with $1.5 < k < 8.5$. For the apex diameter and bias voltage for the tip under discussion (recon 31), we find that $k = 4.0$, i.e., well within the span of validity for the approximation. However, it is instructive to put this into wider context with a few more numerical examples as follows.

Consider the hypothetical situation of the same GaN tip being fully depleted of free carries but without an inversion layer forming in the apex region under the SV considered. In this scenario, the region of the apex behaves as dielectric media (with the same dielectric constant as GaN) and our Poisson simulations yield $E_a$ = 5.6 V/nm and $k$ = 18. As a final example, consider a tip of the same approximate geometry but composed of dielectric, amorphous $SiO_2$ instead of GaN. Then, under the same SV conditions our simulation yields $E_a$ = 0.7 V/nm and $k$ = 90. We point out these comparisons in order to demonstrate that the often-quoted *k*-factor approximation for computing $E_a$ should be used with care since it is only valid under very specialized circumstances and will certainly not generalize to complex, multilayer specimens composed, e.g., of metals, dielectrics, and semiconductors—subjected to various SV conditions that may influence depletion and inversion in the vicinity of the tip apex.

Finally, it is worth pointing out that according to Mancini, et al (Ref. 6), post-ionization theory (Ref. 31) predicts that $E_a$, ~ 24 V/nm should, on average, yield roughly equivalent Ga and N atomic concentrations—thus approximately recovering stoichiometric GaN. Figure 17 (b—e) illustrates that is indeed the case for recon 31. However, Ref. 6 also predicts the average ionic fractions of $Ga^{2+}/Ga^{1+}$ (CSRs) should be within the range of ≈ 0.07—0.1 under these conditions. The spatially-resolved CSRs shown Figs. 17 (b—e) clearly exceed this average, estimated range. Therefore, insofar as our electrostatic simulations are sufficiently accurate in predicting $E_a$, the post-ionization theory as applied by Mancini, et al, yields plausible, average results for CSRs, but does not reveal the finer, spatially resolved effects we observe experimentally.

## 5. Discussion of uncertainties

There are numerous sources of uncertainty that can influence the APT-derived assessment of composition. In Sections 5.1—5.5 we describe what we regard as the more significant of these issues in their order of importance. Some representative examples for the *c*-axis specimens are illustrated using sample h; other *c*-axis samples displayed similar trends and need not be presented in similar detail.

### 5.1 Ambiguous assignments in the range of (14—16) Da

Consider the mass spectrum associated with the TLROI of recon 1 with PE = 2 fJ and hydrogen ignored. The relative counts at 14 Da and 15 Da reasonably support the assignment of $N^{1+}$. However, the relative counts between 14 Da and 14.5 Da also justify the $N_2^{2+}$ assignment. If, however, the (14—15) Da range is assigned entirely to $N^{1+}$, we find the (at. %) composition fractions of N, Ga, and In contained in ROI 1 are 55, 39, and 5.6, respectively. A similar trend is found for ROI 2. Considering recon 4 with PE = 50 fJ, and H again ignored, the relative counts between 14 Da and 15 Da poorly justify an assignment to $N^{1+}$, but the relative counts between 14 Da and 14.5 Da still favor an $N_2^{2+}$ assignment. If we nonetheless assign the (14—15) Da range to $N^{1+}$ in recon 4, we find that the (at. %) composition fractions of N, Ga, and In contained in the corresponding ROI 1 are 46, 48, and 6.3, respectively, and a similar trend if found for ROI 2. Comparing both of these PE cases to their counterparts in Table 3 reveals that such 14-Da assignments necessarily increase the measured In fractions because fewer net counts are assigned to N. Furthermore, over the PE range available, adherence to the convention where 14 Da is assigned to $N^{1+}$ generally results in APT overestimating the In fraction compared to the RBS result. We next discuss how the presence of H can further complicate peak assignments in the (14—16) Da range.

### 5.2 Hydrogen

Hydrogen in an inescapable background contaminant in the APT tool and it can also be present both in and on a specimen tip.[2,3,4] Furthermore, H may also exist as a contaminant in GaN.[43] We generally find that the measured (elemental) H concentration depends upon PE. For example, in ROI 1 of recon 1 (PE = 2 fJ) $H^{1+}$ is observed with a fractional concentration of ≈ 0.04 at.%. In ROI 1 of recon 5 (PE = 50 fJ) $H^{1+}$, $H_2^{1+}$ and $H_3^{1+}$ are all observed and yield a fractional elemental concentration for H of ≈ 0.8 at.%. Of course, the assignment of $H_2^{1+}$ is indistinguishable from possible $He^{2+}$.

Regardless of its actual spatial origin, e.g., background, tip surface adsorbate, photo-desorbed from the chamber walls, or bulk chemical constituent of the tip, hydrogen is mobile in the strong electric field near the tip and it can form complex species with other constituent ions—particularly nitrogen.[3,4] Also, for all of the PE cases above 2 fJ, the relative counts between 14 Da and 15 Da cannot be reconciled by the isotopic ratio provided with a $N^{1+}$ assignment; a preferred alternative choice is to assign 14 Da to $N_2^{2+}$ and 15 Da to $N_2H_2^{2+}$.

Adopting these preferred assignments, the net H concentrations for ROI 1 (PE = 2 fJ) of recon1 and ROI 1 of recon 4 (PE = 50 fJ) becomes ≈ 0.2 at. % and ≈ 0.8 at. %, respectively. At higher PE levels, a 16 Da peak emerges that we assign to $N_2H_4^{2+}$. Therefore, even if we assume all the H originally resides within the specimen tip, inclusion of the H counts will have a negligible on the APT-derived concentration of indium. Additionally, we note that axial concentration profiles for H of the TLROIs generally show a monotonic increase with depth. Such behavior would support the argument that H is primarily a surface contaminant since the tips are conical and the surface area interrogated necessarily increases as ions are field-evaporated away. By contrast, our separate APT studies (not shown) of uniform-diameter GaN nanowires generally reveal TLROI *z*-axial concentration profiles for H that are correspondingly uniform. Finally, we offer the interesting summary observation that while the presence of H has little impact on the uncertainty of the indium concentration measurement, it does offer a means to more justify the argument that peaks in the (14—16) Da range are primarily associated with $N_2$.

5.3 PE-dependent ambiguity for assignment of $^{113}In^{2+}$.

Besides PE-dependent issues associated with N and H, the ranging of the mass-spectral peak at ≈ 56.46 Da also shows a PE-dependent ambiguity. For PE in the range of (2—50) fJ, the isotopic ratio of the $In^{2+}$ peaks favor assigning 56.46 Da to indium. However, for PE > 50 fJ the relative counts between peaks at 55.46 Da and 56.46 Da supports an assignment to $GaN_3^{2+}$. Performing a Saxey-plot analysis may help elucidate these observations.[44] However, such distinctions are academic within the present context since they will have a negligible effect on the assessment of indium concentration at the relative count levels observed. Nonetheless, they are worth pointing out as sources of error that could complicate other IVAS analyses, e.g., assessment of In clustering.[45]

5.4 2D plots of relative Ga density, spatial dependence of Ga CSR, and possibility of FIB-induced Ga implantation

Figure 7 concerns sample h and indirectly reveals the 6-fold rotational symmetry about the *c*-axis expected for the wurtizite structure; it also suggests that there could be an associated 6-fold dependence of the concentration profile that originates from crystal symmetry. However, Figs. 9 (a, b) illustrate that for this relatively low PE case (2 fJ), such an azimuthal dependence of the concentration profile is insignificant. Nonetheless, the radial dependence of the concentration is seen when Ga is decomposed into its three observed charge states. Figs. 10 (a, c) show that the combined $Ga^{2+}$ and $Ga^{3+}$ counts dominate the center of the reconstruction while $Ga^{1+}$ falls to a minimum at the center. This effect is consistent with a purely electrostatic artifact described earlier, and post-ionization theory [Ref. 31], since the vicinity of the tip apex should experience a relatively higher electric field. Indeed, other workers have shown that relatively low PEs, and high electric fields, favor the emission of $Ga^{2+}$ and $Ga^{3+}$ ions.[6] Interestingly, however, for sample h with PE = 2 fJ, the radial dependence of the indium concentration and CSR does not directly mimic that of Ga; Figs. 10 (b, d) illustrate that indium concentration is dominated by combined $In^{2+}$ and $In^{3+}$ counts across the entire diameter of the reconstruction while $In^{1+}$ is a minor contribution. The evolution of these effects for *c*-plane samples as the PE is increased is discussed next.

The radial concentration profiles for Ga and N (with respect to the TLROI) for sample h, where PE = 100 fJ, are illustrated in Figs. 13 (a) and 14 (a). With increasing PE, the expected trend for Ga emerges (Ref. 5) whereby the Ga concentration profile is entirely dominated by $Ga^{1+}$ -- albeit $Ga^{1+}$ shows a relative minimum at the center of the reconstruction. However, the situation for indium is more complicated. As shown in Figs. 13 (b) and 14 (b), the radial concentration profiles of $In^{1+}$ and $In^{2+}$ vary in an opposing sense such that they add up to an approximately uniform indium concertation across the diameter of the reconstruction.

Choice of the 20 nm dia ROI 2 was guided by the observation that over various PE cases for *c*-axis GaN, the axial concentration profiles for Ga and N (in the GaN portions of the recons) are closer to (physically) expected equality than for the associated TLROIs. Examples of such behavior for PE = 2 fJ, 10 fJ, and 100 fJ are shown in Figs. 6 (b), 11 (b), and 12 (b), respectively. However, this trend does not continue into the InGaN regions for all the samples. In particular, for samples h and f (higher indium concentrations) Tables 3 and 4 indicate that for a given PE up to ≈ (50—100) fJ the integrated Ga, N, and In concentrations are roughly equivalent in the corresponding ROI 1 and ROI 2. For sample c (lowest In concentration) the difference between the integrated concentrations between ROIs 1 and 2 emerges for PE ≈ 10 fJ.

The asymmetric, *c*-axis evolution of *m*-plane GaN tips was described in Section 3.1.2 and illustrated in Figs. 16—19. The apparent correlation with growth habit and PEC was also discussed. It is worth noting, however, that even though our electrostatic analysis precluded a direct influence of $\vec{P}_S$ upon $E_a$ in defining how an *m*-plane tip would evolve as ions are field-evaporated away, we point out that Megalini [Ref. 39] attributed the polarity-dependent asymmetry in *c*-axis PEC etching to a mechanism whereby photo-generated holes drift under the action of $\vec{P}_S$. Thus, insofar as laser-assisted field evaporation is analogous to a PEC etch process in defining the evolution of a biased semiconductor APT tip, e.g., photoelectrons pulled to the tip base, photogenerated holes pulled to the tip apex, and the generation and emission of positive ions, the role of spontaneous polarization deserves further study in efforts to identify the precise physical mechanism responsible for the observed asymmetric tip evolution.

Concerns have also been raised regarding compositional errors that could arise from unintentional FIB-induced implantation of Ga ions during tip preparation.[46] However, we find no significant difference in the PE-dependent composition for single GaN nanowires (Refs. 5, 47), which did not require any FIB milling, with trends revealed in the present study. The issue of accidental Ga implantation is likely of lesser concern for analyses within ROIs that omit sections of a tip's surface within the FOV but should be considered carefully if this is not the case.

### 5.5 Dependence of ion counts on ROI placement and depth

Another source of uncertainty is that the detected ion counts within an ROI can depend upon its orientation and placement. Figure 21 (a) illustrates the placement of 3 ROI cylinders that are oriented axially in the TLROI of recon 31. The associated graphs, Figs. 21 (b, c, d), show the depth dependence of the ion counts for Ga, N, and In. The placement of these ROIs is chosen to roughly conform to a similar analysis of *m*-plane GaN as described in Ref. 18. For each case, the total detected counts near the tip apex is greater than the counts for the maximum depth indicated. The effect is most pronounced for ROI 2 of Fig. 21 (b). Also, while both ROI 1 and 2 fall on the "laser facing" side of the TLROI, the total counts associated with ROI 2 are substantially higher.

The spatial dependence of ion counts for a *c*-plane specimen (recon 3) is shown in Fig. 22. Here again, the ROI on the "laser facing" side shows the greater overall counts, however, the ROI on the "laser shadowed" side shows very similar depth-dependent behavior. The coaxially-placed ROI shows the lowest, but most uniform, counts as a function of depth.

The physical origin of these effects is not immediately clear, and we hesitate to speculate on their origin and opt instead to simply report the observations. However, mechanisms involving lattice symmetry, ion trajectories, thermal gradients, and surface diffusion of species, that have been variously described in relation to similar observations, may certainly play a role.[5,7] Whatever their origin, these results show that the assessment of uncertainty can be a rather complex issue. For example, while axial concentration profiles within an ROI may suggest that the ratio of species remains a constant, as is approximately the case shown in Figs. 18 and 19, the absolute number of counts may vary substantially. Hence, the fractional uncertainty of the concentration of a specie may be spatially dependent.

### 5.6 Comments on reconstruction fitting parameters

In generating reconstructions, one is generally compelled to regard DE, ICF, and S/C as non-physical fitting parameters, rather than physical quantities. These parameters are often chosen from experience and intuition in order to force the reconstruction to conform with separately measured features of the specimen tip. Such measured features can include length of material removed in ATP, thickness of layers, and so on. Moreover, one can often generate reconstructions that conform to measured dimensions even if the volume-per-ion, which is input into the IVAS software, is incorrect. Indeed, the software offers default options to automatically choose ion volumetric data regardless of the actual material under consideration. Therefore, solely using APT reconstruction methods to infer reliable structural, crystallographic, and density information should be approached with caution.

## 6. Conclusions

Laser-assisted atom probe tomography (APT) was performed on a series of four, *c*-plane GaN/$In_xGa_{1-x}N$/GaN test structures with $x$ = 0.112, 0.056, 0.034 and 0.030 as separately determined by Rutherford backscattering analysis (RBS). For laser pulse energies (PEs) confined to the approximate range of 2—50 fJ, APT assessment of $x$ conformed to the RBS results to within the estimated RBS uncertainty $\Delta x = \pm 0.01$. Moreover, compared with the integrated concentrations of Ga and N, the APT-determined values for $x$ were relatively insensitive to the sizes of the region-of-interest (ROI) analysis volumes. The deviations from stoichiometry of measured concentrations of nitrogen and group III elements depend on the group III alloy composition. Therefore, the APT tool conditions that reproduce RBS-conforming $x$ values found for samples examined herein will likely not generalize to material containing a higher indium mole fraction.

APT was also performed on *m*-plane GaN/InGaN/GaN core-shell structures. Concentration profiles for $Ga^{2+}$ and $Ga^{1+}$ (in the GaN core region) viewed along [0001] revealed a spatial asymmetry that correlated with a displacement of the tip apex in the $[000\bar{1}]$ direction as ions are field-evaporated away. When viewed along $[\bar{1}2\bar{1}0]$, both the tip apex and concentration profiles for $Ga^{2+}$ and $Ga^{1+}$were essentially symmetric. The symmetry of the APT tip evolution was found to correlate with both the growth habit of *m*-axis GaN nanostructures, and the anisotropic photo-electrochemical (PEC) etching of *m*-plane GaN.

Electrostatic simulations were performed on realistic cases of *n*-type GaN tip-electrode geometry and tip-electrode bias. The results revealed that formation of a *p*-type inversion layer over the tip apex precluded the possibility that the apex electric field could be transversely perturbed by the built-in spontaneous polarization, since the polarization induced surface charge would be screened by the free holes in the inversion layer. More broadly, our simulations also reveal that for APT analysis of semiconductors for cases where a conducting inversion layer is present, the apex field may be estimated using the simplified *k*-factor approximation—which was originally derived for metal tips. However, it should be borne in mind that the *k*-factor approximation fails for purely dielectric tips. Therefore, care must be taken in treating the electrostatics of complex specimens that may be variously composed of metals, semiconductors, and dielectrics.

Analysis of both *c*-plane and *m*-plane specimens revealed variations in detected counts (for all species) which depended upon ROI placement and analysis depth. This latter effect can produce spatially dependent variations in the computed error, and fractional uncertainties of detected species, which are based on counting statistics. Therefore, these issues should be considered when in quoting uncertainties in APT results.

Acknowledgments:

The authors gratefully acknowledge insightful discussions with A. N. Chiaramonti, K. A. Bertness, D. R. Diercks, and B. L. Gorman.

The National Institute of Standards and Technology (NIST) supported the contributions of NIST-affiliated authors.

Authors affiliated with the University of New Mexico acknowledge that the nanostructure growth work was supported by the Defense Advanced Research Projects Agency (DARPA) under award number D13AP00055 and by the NSF under cooperative agreement EEC-0812056. Any opinions, findings, and conclusions or recommendations expressed in this material are those of the author(s) and do not necessarily reflect the views of DARPA or the National Science Foundation.

## Tables and Figures

| sample type | In concentration (RBS) | | layer thickness | | |
|---|---|---|---|---|---|
| | at.% | *x* | *L* | $L_c$ | Measurement method |
| h | 5.60 | 0.112 | 55.4 | 13.2 | XRD |
| f | 2.8 | 0.056 | 360 | 23.0 | |
| d | 1.7 | 0.034 | 327 | 20.0 | |
| c | 1.5 | 0.030 | 330 | 20.0 | Estimated from growth rate |

Table 1. Summary of *c*-plane $In_xGa_{1-x}N$ test structures by sample type. The RBS-determined indium concentration is reported in at.%, with an estimated uncertainty of ± 0.5 at.%. Indium concentrations expressed in at.% are multiplied by 0.02 to convert to mole fraction *x*.
The thicknesses of the $In_xGa_{1-x}N$ layer and capping layer are labeled *L* and $L_c$, respectively.

| charge state | | | | comments |
|---|---|---|---|---|
| | $1^+$ | $2^+$ | $3^+$ | |
| specie | ranged peaks (Da) | | | |
| Ga | 68.976, 70.925 | 34.463, 35.462 | 22.975, 23.642 | Volume per ion in reconstruction is 0.0114 $nm^3$ (as for GaN). Correcting for true InGaN volume is insignificant. |
| N | 14.003 | 7.0015 | | |
| In | 112.90, 114.90 | 56.450, 57.450 | 38.300 | |
| $N_2$ | 28.006 | 14.003 | | |
| $N_2H_2$ | | 15.011 | | Ranged as $N_2$ since H contributes no volume. |
| $N_2H_4$ | | 16.019 | | |
| $N_3$ | 42.009 | | | |
| GaN | | 41.464, 42.464 | | |
| $GaN_2$ | | 48.466, 49.465 | | Prominent when PE > 1 pJ. |
| $GaN_3$ | | 55.467, 56.467 | | |
| H | | | | Assigned but not ranged. Ambiguities exist with possible He. |
| $H_2$ | | | | |
| $H_3$ | | | | Assigned but not ranged. |

Table 2. Typical ranged species, associated charge states observed, and comments related to reconstructions.

| recon | | ROI 1 | | | | ROI 2 | | | DR | DE | ICF | $L_R$ | S/C | ID |
|---|---|---|---|---|---|---|---|---|---|---|---|---|---|---|
| | T | PE | N | Ga | In | N | Ga | In | | | | | | |
| 1 | 54 | 2 | 58 | 37 | 5.2 | 58 | 37 | 5.2 | 0.4 | 0.19 | 2.0 | 52 | 1.12 | 1-6_01256v21 |
| 2 | | 5 | 57 | 38 | 5.0 | 56 | 39 | 5.1 | | 0.32 | | | 1.00 | 2-1_01257v10 |
| 3 | | 10 | 55 | 40 | 5.3 | 55 | 40 | 5.4 | | 0.32 | | | 1.02 | 2-2_01258v09 |
| 4 | | 50 | 48 | 46 | 6.0 | 51 | 43 | 5.8 | | 0.26 | | | 1.20 | 2-3_01259v06 |
| 5 | | 100 | 43 | 51 | 6.0 | 48 | 46 | 5.9 | | 0.28 | | | 1.14 | 6-2_01020v07 |
| 6 | | 1000 | 16 | 76 | 8.4 | 19 | 72 | 9.1 | | 0.24 | | | 1.20 | 5-2_01030v06 |
| 7 | 26 | 2 | 58 | 37 | 4.8 | 57 | 38 | 5.1 | 0.4 | 0.29 | 2.0 | 52 | 1.10 | 1-2_01252v18 |
| 8 | | 5 | 56 | 39 | 4.9 | 55 | 40 | 5.1 | | 0.27 | | | 1.13 | 1-3_01253v12 |
| 9 | | 10 | 55 | 40 | 5.4 | 55 | 40 | 5.8 | | 0.29 | | | 1.12 | 1-4_01254v05 |
| 10 | | 50 | 48 | 46 | 5.8 | 52 | 42 | 6.2 | | 0.23 | | | 1.00 | 1-5_01255v09 |

Table 3. Data acquisition and analysis parameters per reconstruction for specimens from sample h. The heading of the column enumerating individual reconstructions is abbreviated as, "recon." ROI 1 is a cylindrical region that contains all of the InGaN portion of the TLROI but excludes the top and bottom GaN layers. ROI 2 is a 20-nm-diameter cylindrical subset of ROI 1 that is placed coaxial with the TLROI. Note that $L_R$ (nm) is the length of both ROIs and is set ≈ 2 nm less than the XRD-determined InGaN layer thickness in order to accommodate irregularities and tilts of the GaN/InGaN interfaces. N, Ga, and In enumerate the respective elemental concentrations (at.%) within each ROI. ID lists verbose identifications of each reconstruction to assist in future reference (e.g. for recon 1, "1-6" is the tip number, "01256" is the datafile number, "v21" is the version of the recon). Other abbreviations are given in the text.

| recon | | | ROI 1 | | | ROI 2 | | | DR | DE | ICF | $L_R$ | S/C | ID |
|---|---|---|---|---|---|---|---|---|---|---|---|---|---|---|
| | T | PE | N | Ga | In | N | Ga | In | | | | | | |
| 11 | 54 | 2 | 57 | 40 | 2.9 | 57 | 40 | 3.2 | 0.4 | 0.3 | 2.0 | 40 | 1.08 | 6-2_01208v04 |
| 12 | | 5 | 56 | 41 | 2.8 | 56 | 41 | 2.8 | | | | 50 | 1.15 | 6-3_01211v04 |
| 13 | | 10 | 54 | 43 | 3.2 | 54 | 43 | 3.5 | | | | 42 | 1.10 | 6-5_01216v02 |
| 14 | | 50 | 47 | 50 | 3.7 | 49 | 47 | 3.7 | | | | 26 | 1.09 | 6-6_01215v03 |
| 15 | | 100 | 39 | 57 | 3.7 | 45 | 51 | 3.6 | | | | 41 | 1.11 | 5-6_01217v03 |
| 16 | | 1000 | 9 | 86 | 5.6 | 10 | 85 | 5.7 | | | | 35 | 1.04 | 5-5_01218v02 |

Table 4. Data acquisition and analysis parameters per reconstruction (recon) for specimens from sample f. Both ROI cylinders are placed just below the upper GaN/InGaN interface, are coaxial with the TLROI, encompass only the InGaN layer, and are of length $L_R$ per recon. Comparative FESEM imaging of the tip before and after L-APT data acquisition are used in setting the scale for $L_R$; the scale is not determined from the XRD-measured thickness of the InGaN layer. ROI 1 encloses the TLROI to the length indicated and ROI 2 is 20 nm in diameter. Abbreviations are consistent with Table 3.

| recon | | ROI 1 | | | | ROI 2 | | | DR | DE | ICF | S/C | $L_R$ | ID |
|---|---|---|---|---|---|---|---|---|---|---|---|---|---|---|
| | T | PE | N | Ga | In | N | Ga | In | | | | | | |
| 17 | 54 | 2 | 56 | 42 | 1.5 | 54 | 44 | 1.2 | 0.4 | 0.27 | 2.0 | 1.00 | 155 | 1-1_01093v03 |
| 18 | | 10 | 50 | 48 | 1.6 | 51 | 48 | 1.7 | | 0.27 | | 1.16 | 220 | 1-2_01096v01 |
| 19 | | 100 | 33 | 64 | 2.2 | 44 | 54 | 2.0 | | 0.28 | | 1.07 | 128 | 1-3_01097v04 |
| 20 | | 1000 | 6.1 | 91 | 2.7 | 11 | 87 | 2.8 | | 0.20 | | 1.30 | 133 | 1-4_01098v04 |

Table 5. Data acquisition and analysis parameters per reconstruction (recon) for specimens from sample d. Both ROI cylinders are placed just below the upper GaN/InGaN interface, are coaxial with the TLROI, encompass only the InGaN layer, and are of length $L_R$ per recon. The scale for $L_R$ is determined as described in Table 4. Abbreviations are consistent with Table 3.

| recon | | ROI 1 | | | | ROI 2 | | | DR | DE | ICF | S/C | $L_R$ | ID |
|---|---|---|---|---|---|---|---|---|---|---|---|---|---|---|
| | T | PE | N | Ga | In | N | Ga | In | | | | | | |
| 22 | 54 | 2 | 57 | 42 | 1.4 | 56 | 42 | 1.5 | 0.4 | 0.22 | 2.00 | 1.13 | 62 | 1-1_01220v03 |
| 23 | | 5 | 52 | 46 | 1.5 | 53 | 45 | 1.6 | | 0.28 | 2.00 | 1.22 | 37 | 1-1_01221v02 |
| 24 | | 10 | 47 | 51 | 1.5 | 52 | 46 | 1.6 | | 0.29 | 2.00 | 1.11 | 38 | 1-3_01222v04 |
| 25 | | 50 | 34 | 65 | 1.8 | 48 | 51 | 1.5 | | 0.20 | 2.00 | 1.23 | 44 | 1-4_01223v02 |
| 26 | 26 | 2 | 55 | 44 | 1.4 | 54 | 44 | 1.6 | 0.4 | 0.28 | 2.00 | 1.10 | 65 | 1-3_01227v03 |
| 27 | | 5 | 52 | 47 | 1.5 | 53 | 46 | 1.7 | | 0.21 | 2.00 | 1.09 | 50 | 2-2_01226v02 |
| 28 | | 10 | 49 | 50 | 1.4 | 51 | 47 | 1.5 | | 0.25 | 2.00 | 1.12 | 45 | 2-1_01225v02 |
| 29 | | 50 | 37 | 61 | 2.0 | 47 | 51 | 1.7 | | 0.29 | 2.00 | 1.17 | 46 | 1-6_01224v04 |

Table 6. Data acquisition and analysis parameters per recon for specimens from sample c. Both ROI cylinders are placed just below the upper GaN/InGaN interface, are coaxial with the TLROI, encompass only the InGaN layer, and are of length $L_R$ per recon. The scale for $L_R$ is determined as described in Table 4. Abbreviations are consistent with Table 3.

| recon | T | PE | QWs in recon | DR | DE | ICF | S/C | $L_R$ | ID and notes |
|---|---|---|---|---|---|---|---|---|---|
| 30 | 54 | 10 | 1, 2, 3 | 0.3 | 0.46 | 2.0 | 1.06 | 80 | 2-2_02097v07 (QWs tilted in recon) |
| 31 | | 15 | 2, 3 | 0.4 | 0.50 | 2.0 | 1.20 | 40 | 1-3_01468v02 (QWs tilted in recon) |
| 32 | | 20 | 2, 3 | 0.3 | 0.25 | 2.3 | 1.40 | 90 | 1-2_01465v03 (QWs show less tilt) |
| 33 | | 10 | no QWs present | 0.3 | 0.20 | 2.0 | 1.10 | 120 | 2-3_02095v03 (GaN only) |

Table 7. Data acquisition and analysis parameters per recon for specimens from *m*-plane samples. As described in the text, the GaN capping layer was comparatively thin and therefore the topmost QW (QW 1) was often consumed in the process of tip alignment. QW 1 was thus unavailable in recons 31 and 32. For the case of these *m*-plane samples, $L_R$ refers to the length of the TLROI and the scale is set by forcing the recons to simultaneously render the QWs as approximately planar and separated by the TEM-determined spacing. Hence, $L_R$ does not typically conform to the true total length of material removed from the tip—as measured by FESEM before and after data acquisition. Due to FIB milling errors, recon 33 has no QWs present but is included to illustrate consistency of the asymmetric tip evolution artifact found in *m*-plane specimens. For recon 33, $L_R$ is estimated by comparing FESEM images recorded before and after APT.

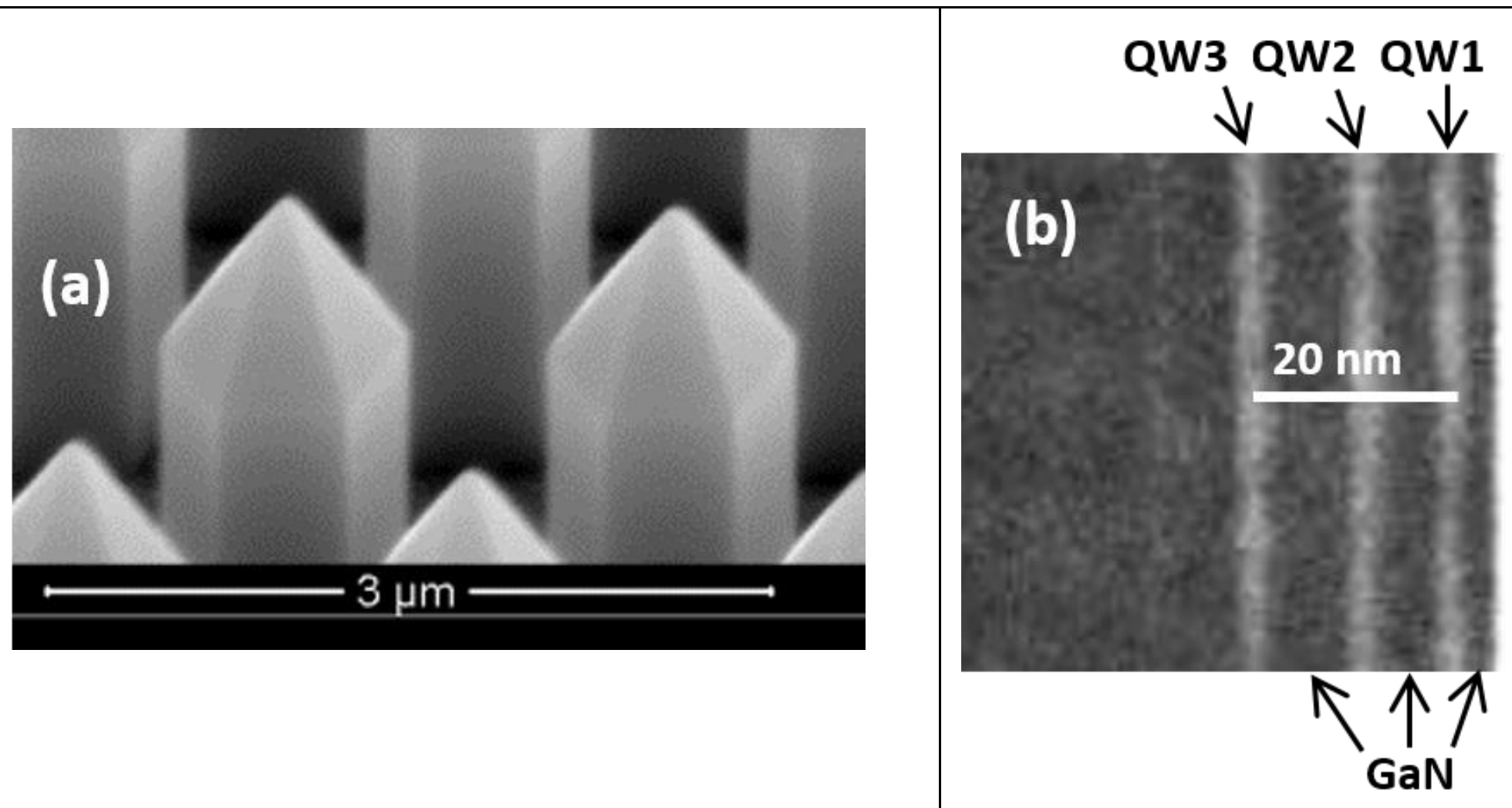

Fig. 1. (a) FESEM image of GaN/InGaN core-shell microposts. The vertical sidewalls conform to the $\{10\bar{1}0\}$ family of *m*-planes. The Ga-polar apex points in the $[0001]$ direction (+*c*). (b) TEM cross section of the core-shell quantum wells (QWs). The QWs reside on *m*-planes and QW1 is nearest the surface. The GaN capping layer is indicated to the extreme right and the two GaN barrier layers reside between the QWs.

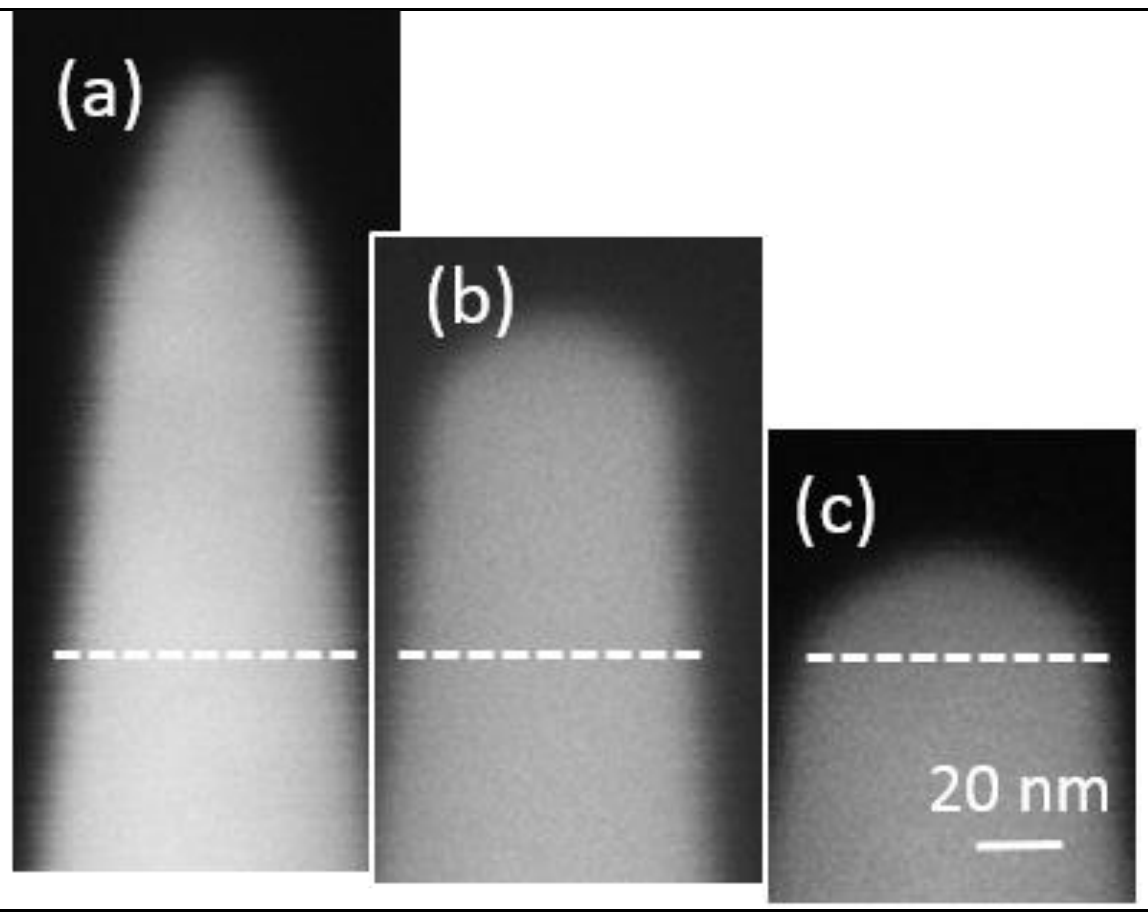

Fig. 2. FESEM images showing the progression of tip evolution for the specimen used in recon 1. (a) Tip after FIB processing completed. The tip apex is composed of residual Ni (from the FIB preparation process) that is ≈ 20 nm thick. (b) Remaining tip after all of the Ni and a portion of the GaN capping layer has been removed by APT. (c) Remaining tip after the remaining GaN capping layer, all of the InGaN layer, and a portion of the GaN template have been removed by APT. In (a) the tip tilts toward the observer and stands 54° from the horizontal; in (b) and (c) the tip stands upright at an angle of 90° from the horizontal.

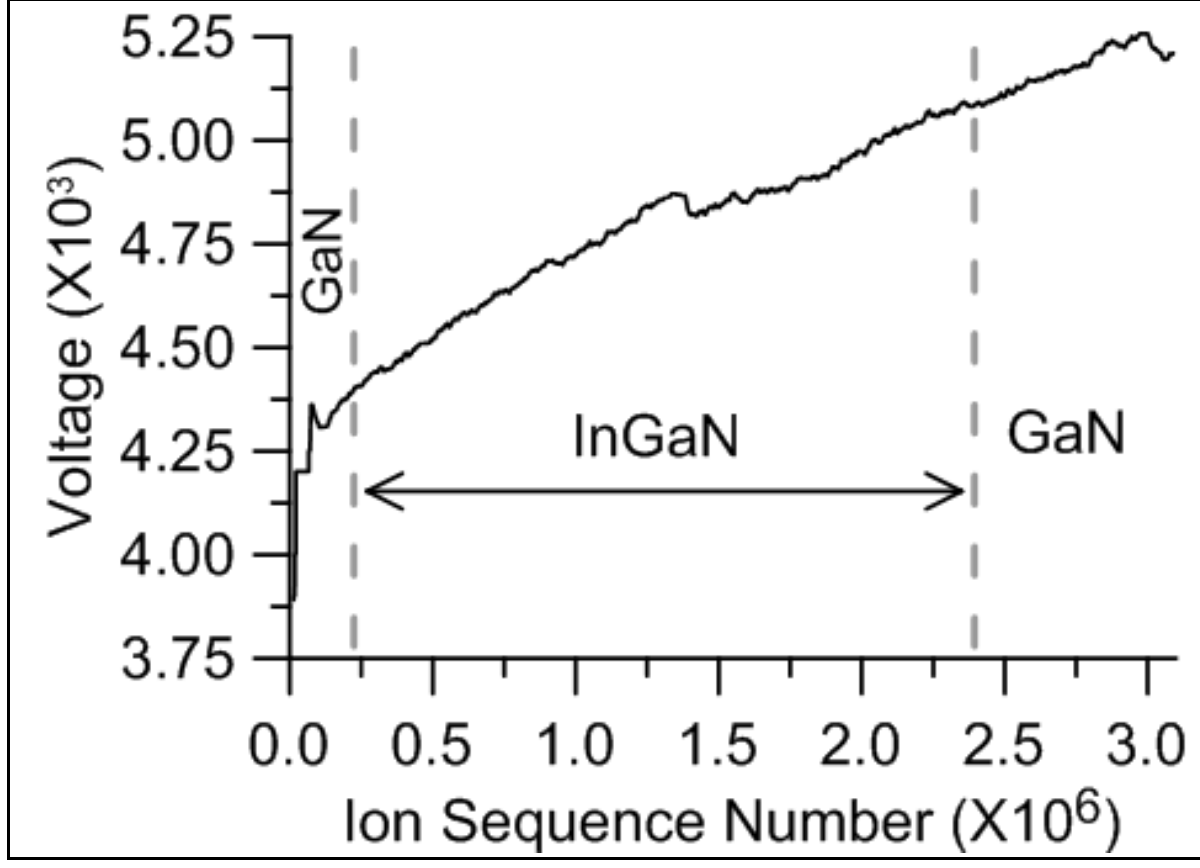

Fig. 3. Variation in specimen voltage for the case of recon 1. The data acquisition starts with the status of the specimen tip as shown in Fig. 1(b) and ends with the status of the tip as shown in Fig. 1(c).

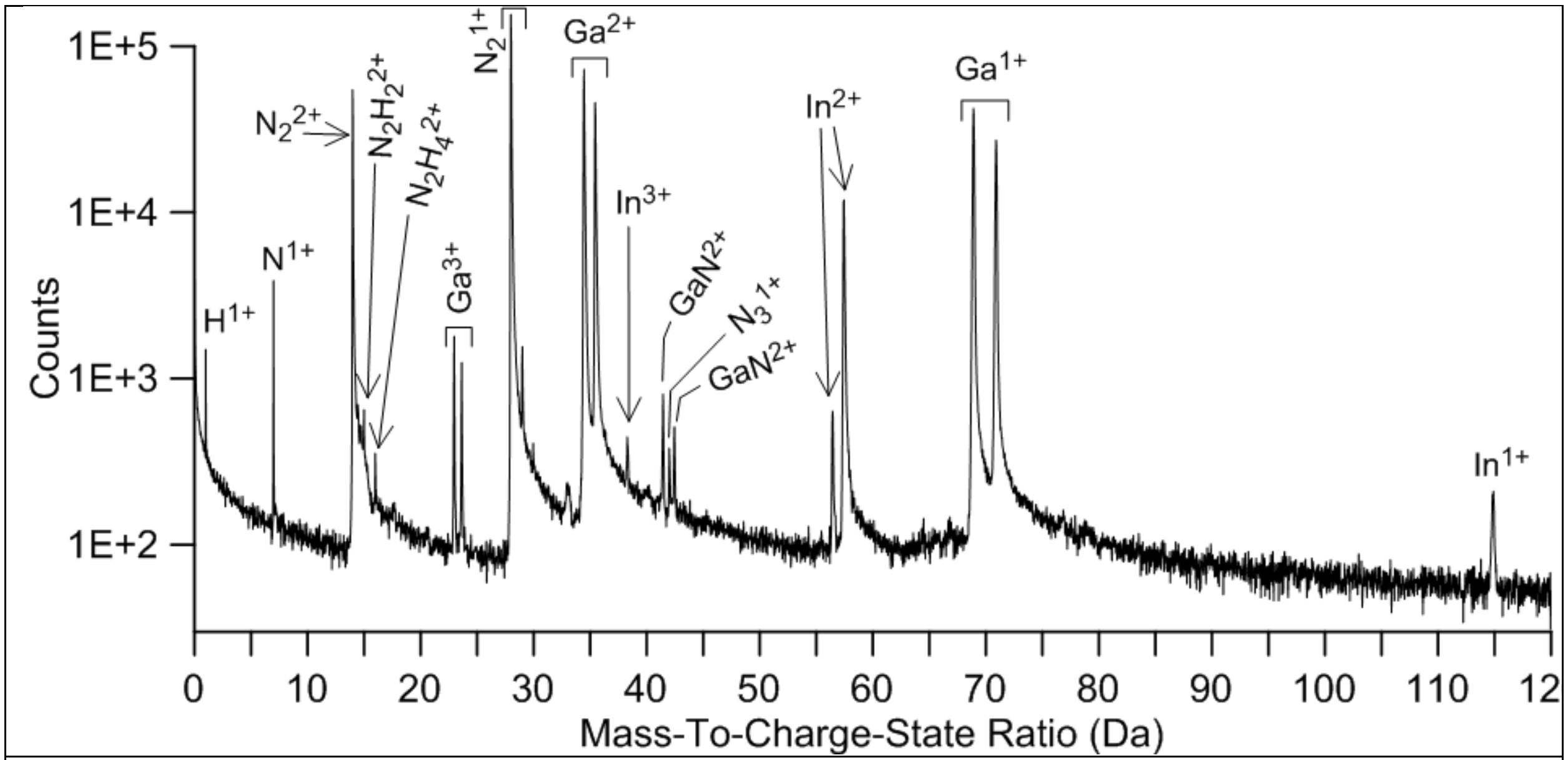

Fig. 4. Full accumulated mass spectrum corresponding to recon 1.
Note that with the $GaN_3$ assignments the $GaN_3$ peak primarily located at 55.467 Da does not overlap with $In^{2+}$, but the $GaN_3^{2+}$ peak at 56.457 does.

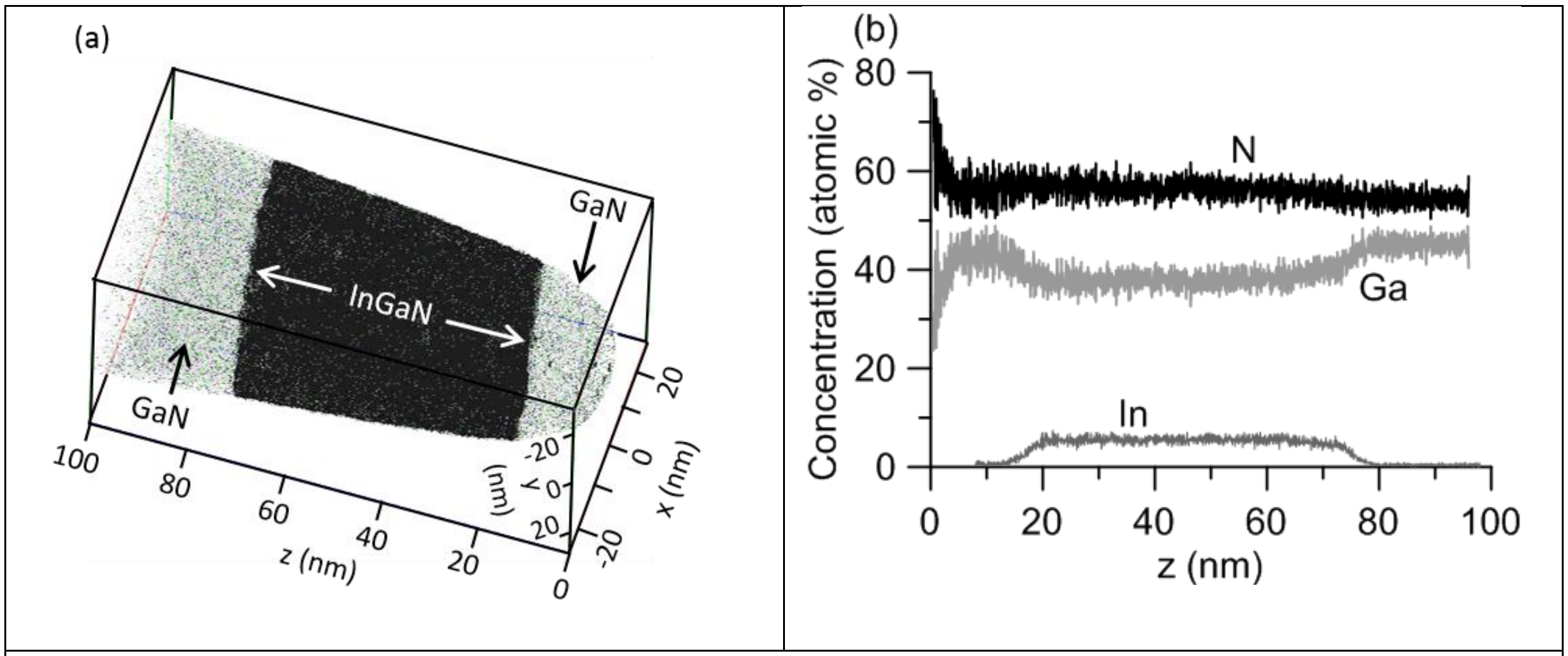

Fig. 5. Recon 1. (a) The InGaN layer, a remnant of the GaN capping layer, and a portion of the GaN template are shown in this TLROI. Due to unavoidable FIB processing issues, the long axis of the tip is tilted by several degrees with respect to the normal of the GaN/InGaN interfaces. The coordinate system indicated preserves the IVAS-defined convention. (b) Axial concentration profile (along z) for In, Ga, and N in the TLROI of recon 1. Note that the first several nm of measured In concentration are omitted since In is insignificantly present in the capping layer and the likely occurrence of $GaN_3^{(2+)}$ will mimic $In^{2+}$. At the 2 fJ PE used for this case, the measured stoichiometry of the GaN template region is (non-physically) N-rich.

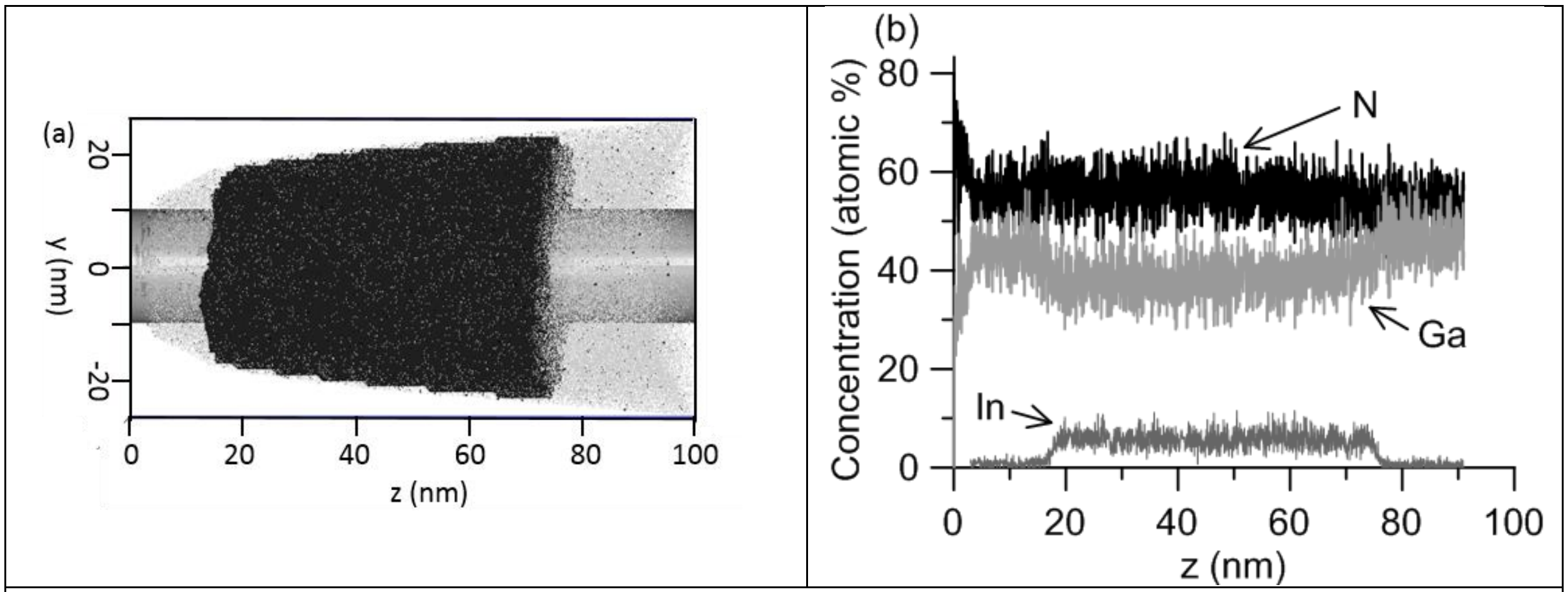

Fig. 6. (a) A 20-nm-diameter cylindrical ROI aligned coaxially with the TLROI of recon 1. The z-axis of the ROI is collinear with the z-axis of the TLROI. (b) Axial concentration profile (along z) for In, Ga, and N in the cylindrical ROI shown in (a). Within the confines of the ROI, the measured N-rich stoichiometry of the GaN template layer is less pronounced than in Fig. 5(b).

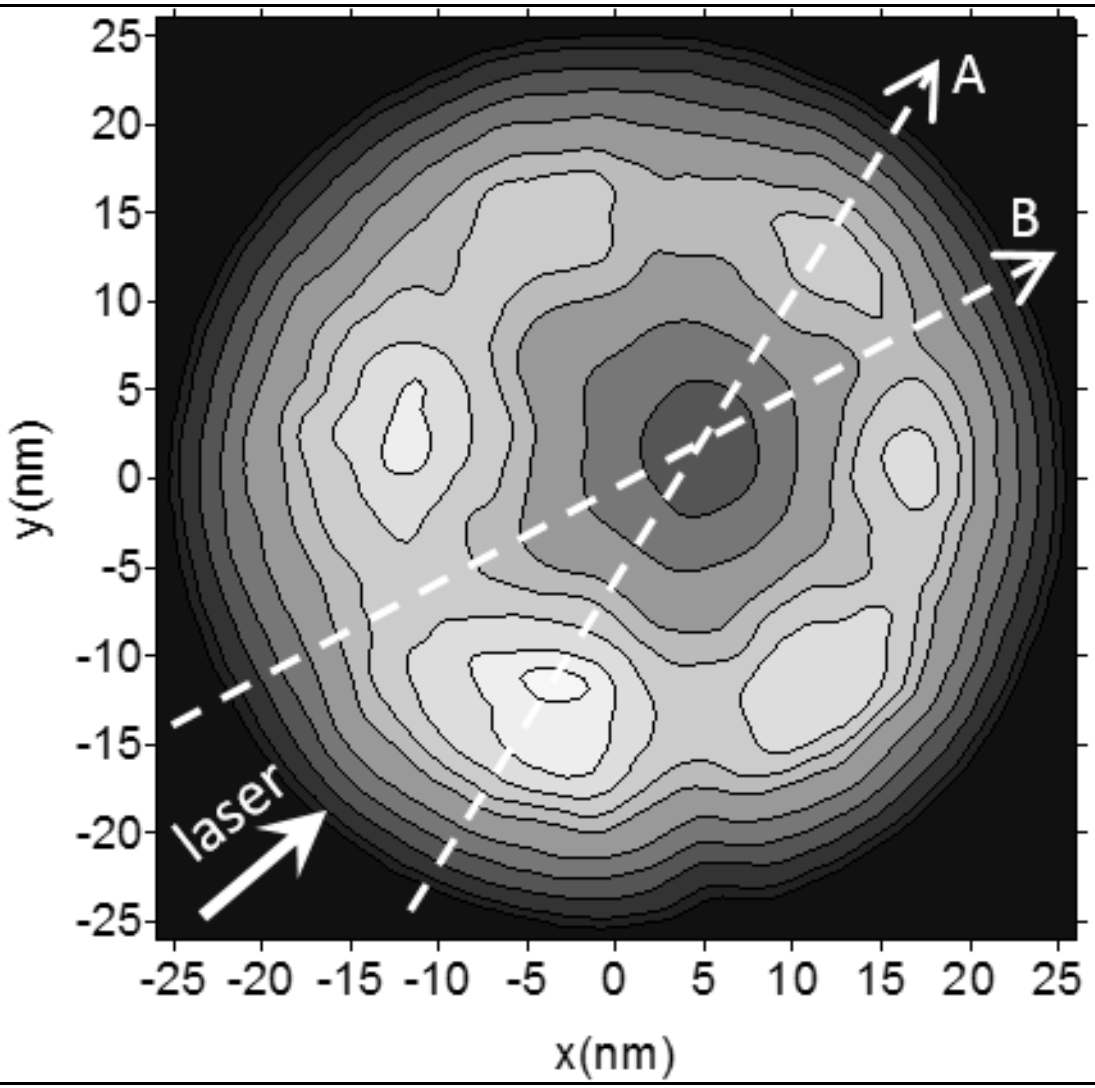

Fig. 7. 2D relative density plot for Ga as projected onto the x-y plane of recon 1. White indicates relative maximum Ga density and black indicates zero density. This map serves as a guide with which to orient ROIs as discussed in the following figures. In the coordinate convention shown, the laser is incident from the lower left and the plot illustrates a 6-fold rotational symmetry similar to prior work.[7] Dashed lines A and B indicate the orientation of respective ROIs that are further illustrated in Fig. 8 with A intersecting regions of relatively high Ga density and B intersecting regions of relatively low Ga density. The displacement of the center of the plot in the positive x-y direction arises from the accidental tilt of the tip in FIB mounting.

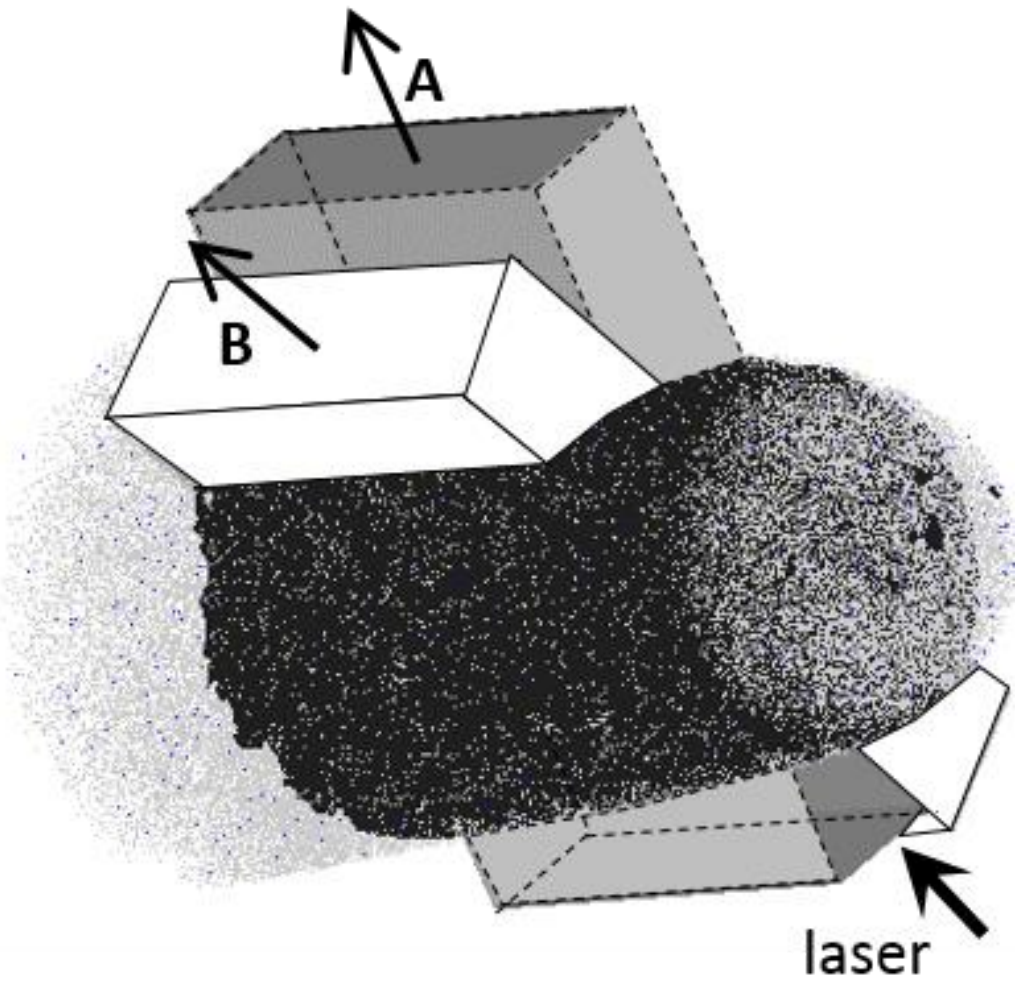

Fig. 8. The InGaN portion of recon 1 intersected by ROIs A and B, as identified by their respective long axes as shown. Both A and B are cuboids and both have dimensions 75 nm × 45 nm × 15 nm. In this view, the laser is incident from the lower right.

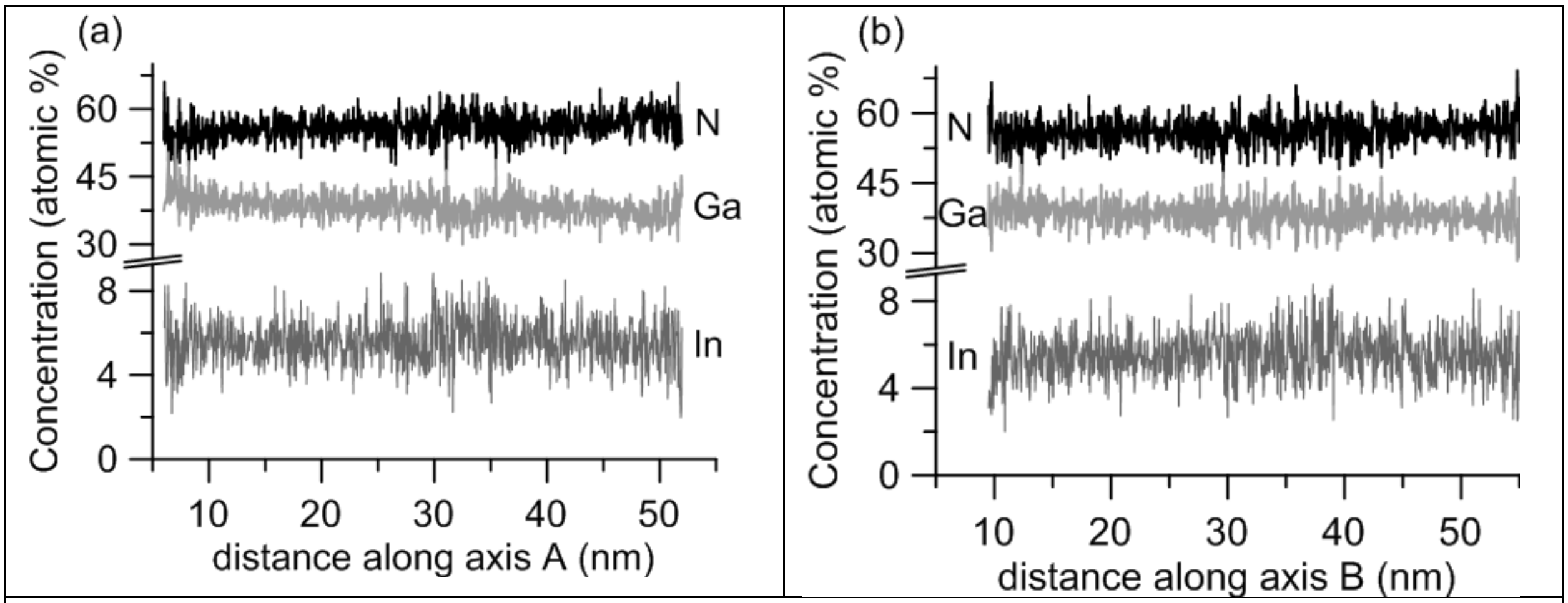

Fig. 9. Axial concentration profiles for N, Ga, and In confined to ROIs A and B as shown in Fig. 8. The respective concentration profiles contained in ROI A and B are illustrated in (a) and (b) above. The results indicate that for recon 1 with PE = 2 fJ, the concentration profiles are nearly constant and essentially independent of the azimuthal placement of the ROIs.

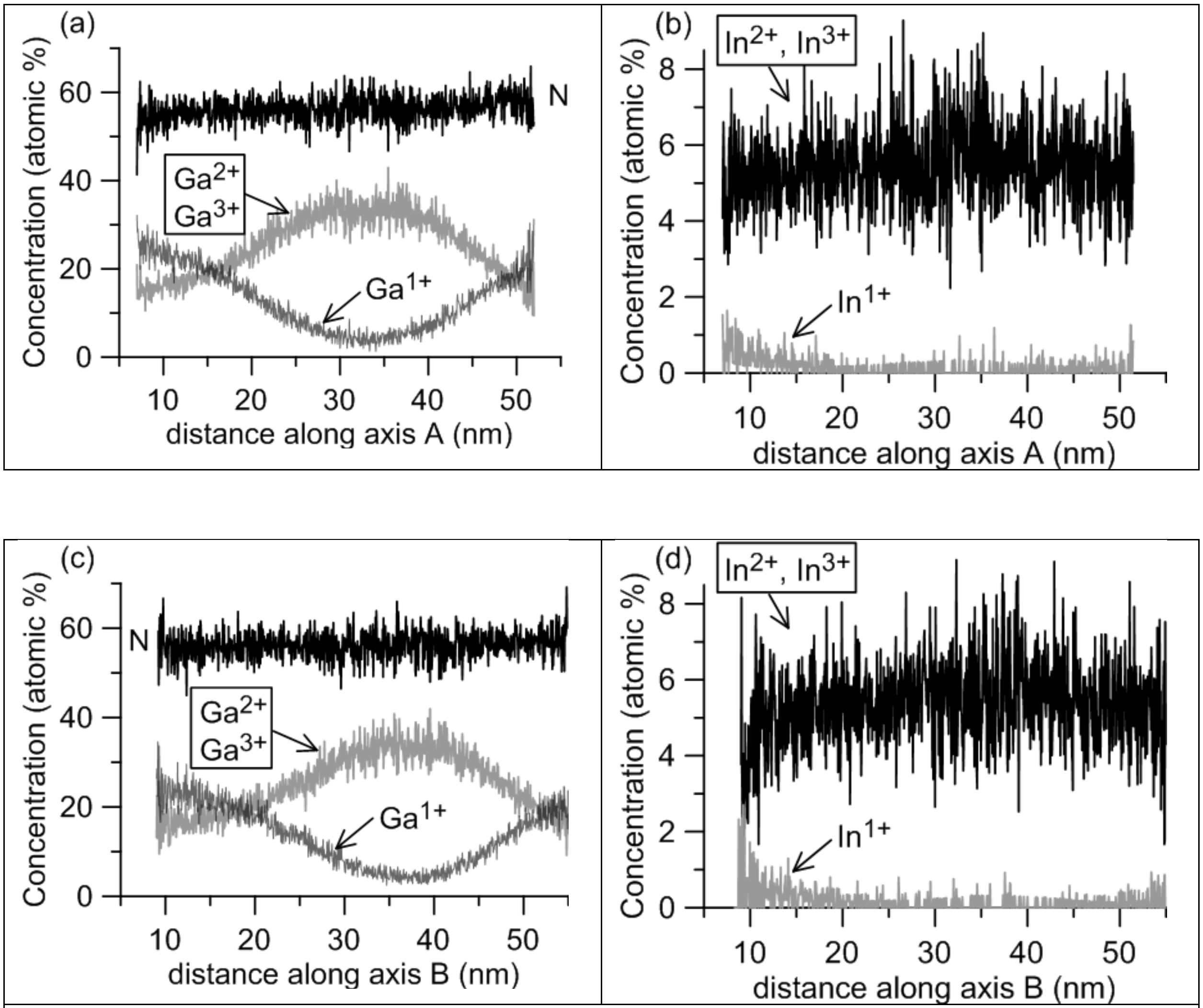

Fig. 10. Concentration profiles in ROIs A and B of Fig. 8 for various charge states of Ga and In. Again, PE = 2 fJ. The bin width per depth increment is 0.05 nm for all cases. (a) The radial dependence of the Ga charge states in ROI A reveals that the combined $Ga^{2+}$ and $Ga^{3+}$ concentrations attain a maximum near the center of the reconstruction with the $Ga^{1+}$ concentration displaying the opposite spatial behavior. Interestingly, the radial dependence of the sum of all 3 Ga charge states results in the nearly uniform concentration profile for Ga as illustrated in Fig. 9 (a). (b) Concentration profiles for the 3 observed charge states of In in ROI A showing that $In^{1+}$ is essentially negligible and the combined concentrations of $In^{2+}$ and $In^{3+}$ show relatively weak radial dependence. (c) Similar to (a) but confined to ROI B. (d) Similar to (b) but confined to ROI B. The rightward displacement of the $Ga^{1+2+,3+}$ extrema in (a) and (c) arise from the accidental tilt of the tip in FIB mounting.

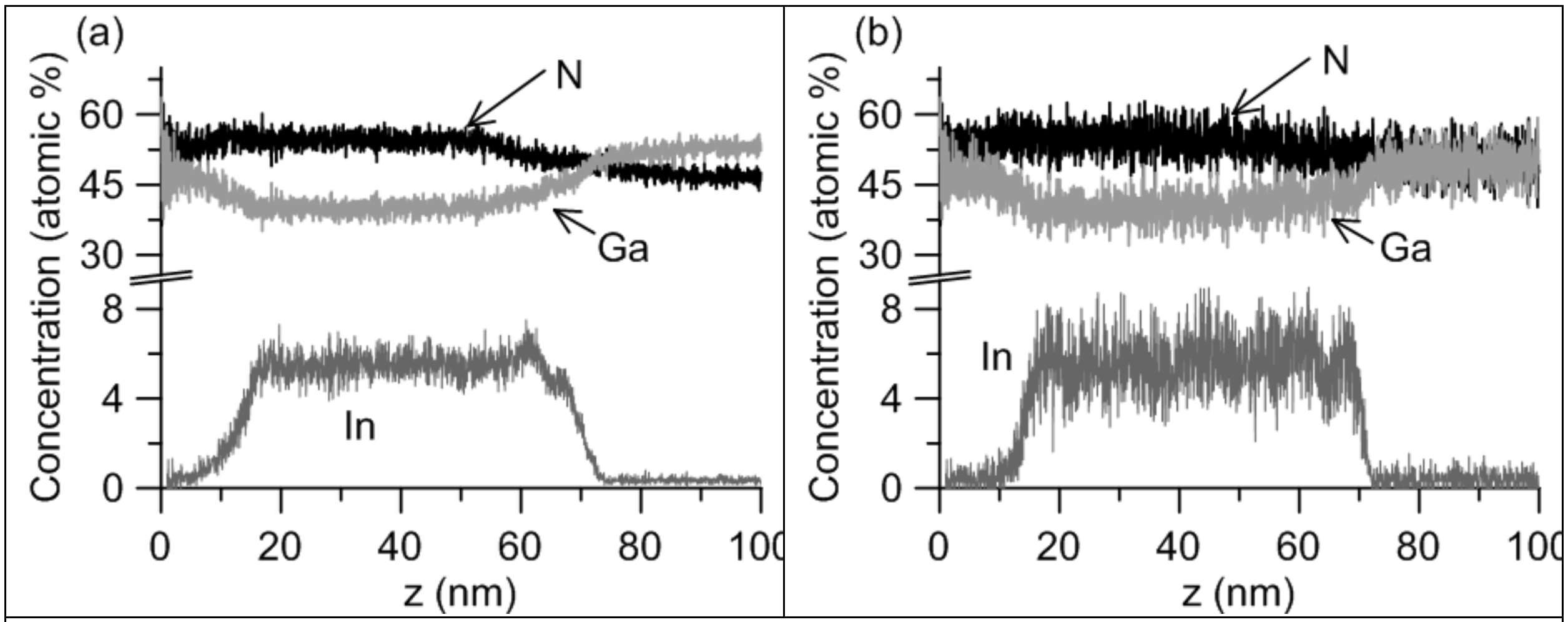

Fig. 11. Axial concentration profiles of N, Ga and In for recon 3 with PE = 10 fJ. The bin width per depth increment is 0.05 nm in all cases. (a) Concentration profiles within full TLROI showing measured Ga-rich composition within the lower GaN layer. (b) Concentration profiles confined to a 20 nm diameter ROI cylinder placed coaxially with the TLROI. In this case the GaN composition appears roughly stoichiometric. Both (a) and (b) indicate non-physical variations in N concentration in going between the InGaN and GaN layers.

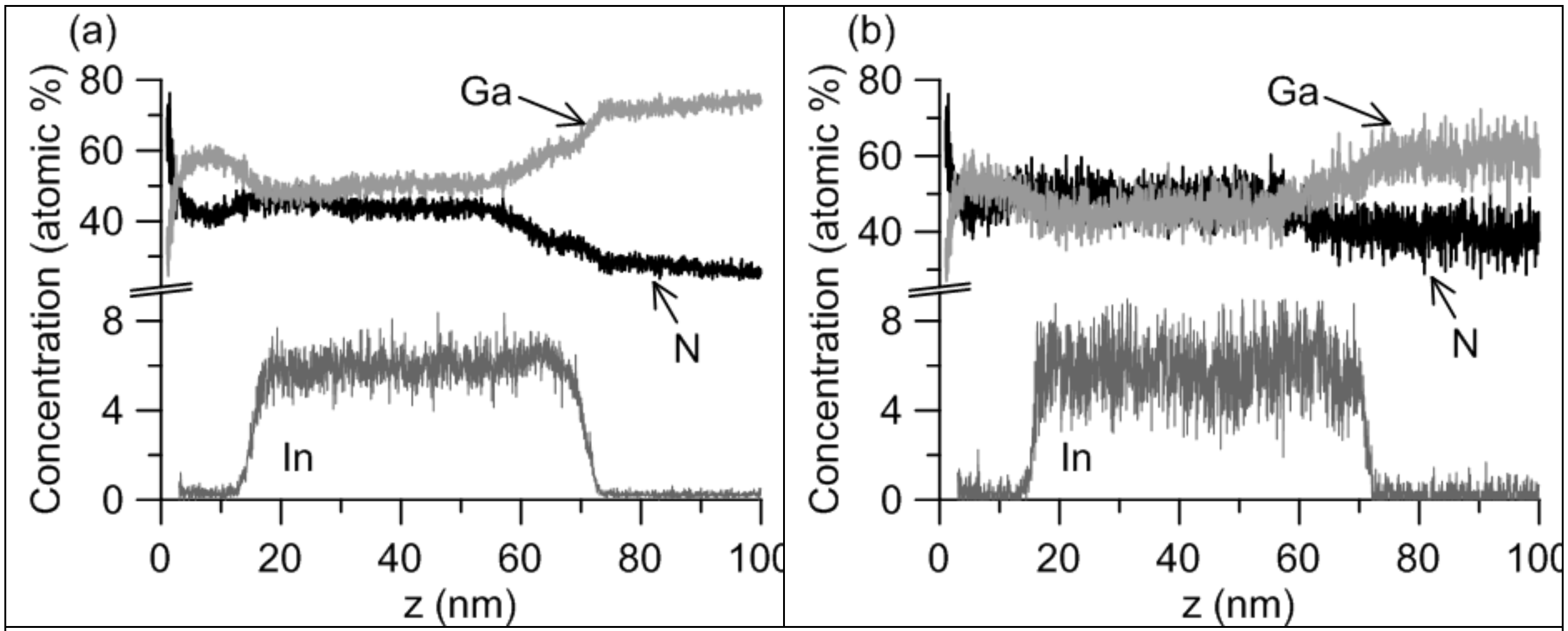

Fig. 12. Axial concentration profiles of N, Ga and In for recon 5 with PE = 100 fJ. The bin width per depth increment is 0.05 nm in all cases. (a) Concentration profiles within full TLROI showing measured Ga-rich composition within the GaN layers. (b) Concentration profiles confined to a 20 nm diameter ROI cylinder placed coaxially with the TLROI. In this case the composition also appears Ga-rich in the GaN layers. Both (a) and (b) indicate non-physical variations in N concentration in going between the InGaN and GaN layers.

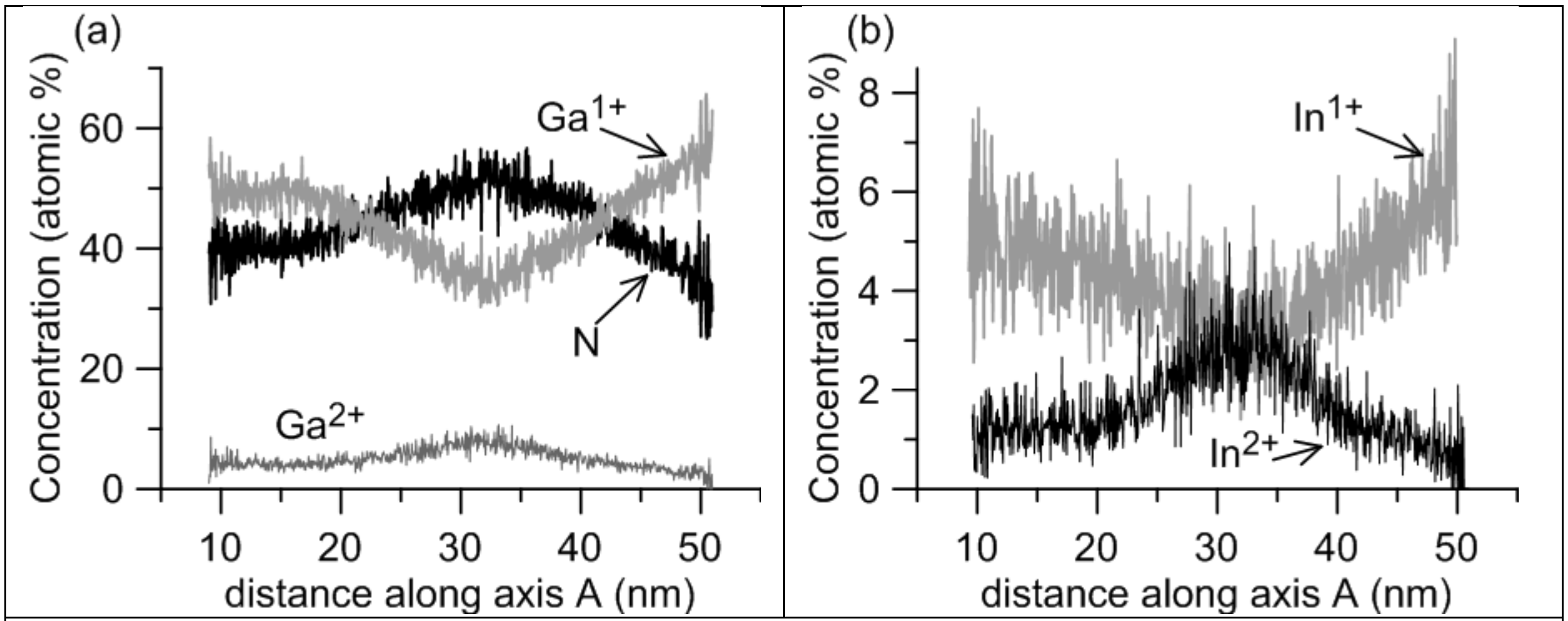

Fig. 13. Axial concentration profiles for N, Ga, and In for ROI A placed into the InGaN portion of recon 5 where PE = 100 fJ. Here, the ROI has the same dimensions and orientation in the TLROI as was described for the eponymous ROI of Fig. 8. However, in contrast to the example shown in Fig. 7, the present case with PE= 100 fJ revealed only approximate radial symmetry of the measured Ga density; no 6-fold azimuthal symmetry was observed. Therefore, to maintain consistency with Figs. 7 and 8, ROIs A and B (see Fig. 14) are oriented with respect to the laser direction and not to the relative Ga density. (a) Concentration profiles for $Ga^{1+}$, $Ga^{2+}$, and N. In this case, both the N and $Ga^{2+}$ concentrations peak at the center of the ROI, but $Ga^{2+}$ is a minor contribution the total Ga concentration. $Ga^{3+}$ is negligible and not shown. (b) Concentration profiles of $In^{1+}$ and $In^{2+}$ illustrate that $In^{2+}$ peaks in the center of the ROI and the summed concentrations result in an approximately constant total indium concentration within the ROI.

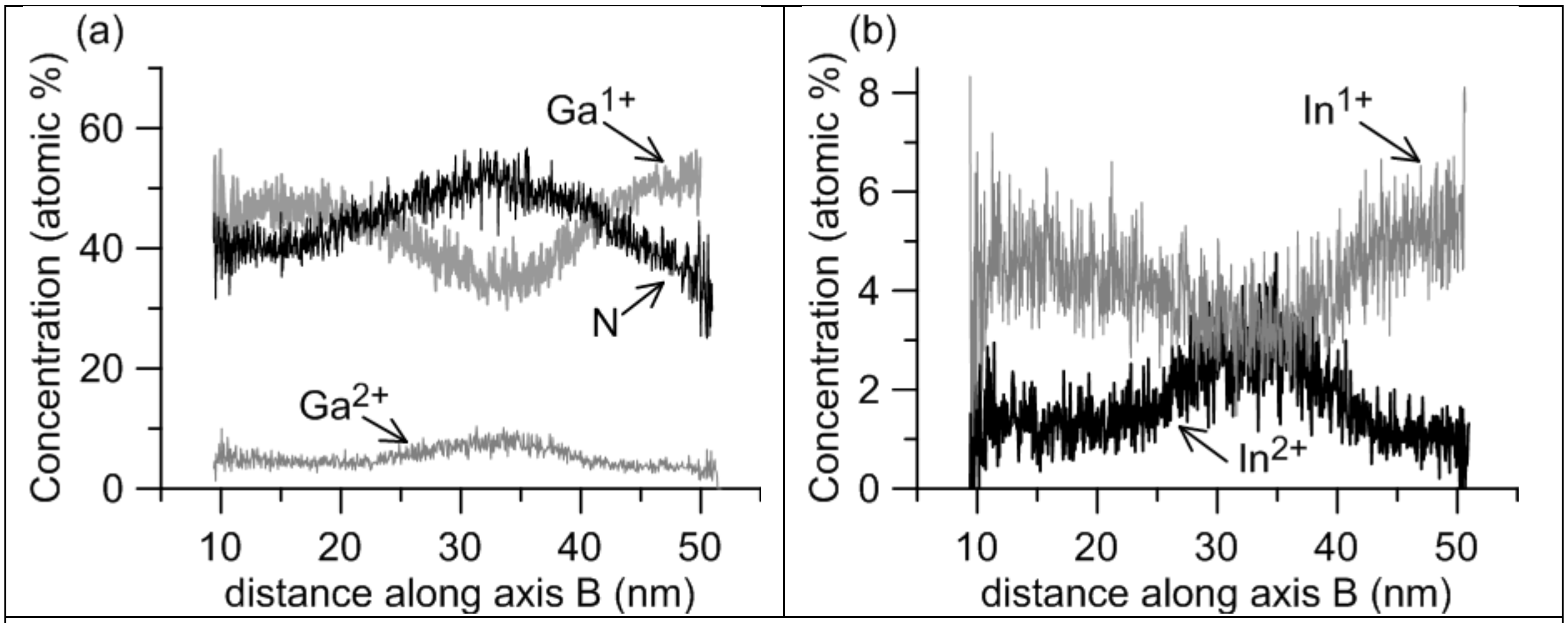

Fig. 14. Axial concentration profiles for N, Ga, and In for ROI B placed into the InGaN portion of recon 5 where PE = 100 fJ. The ROI has the same dimensions and orientation in the TLROI as was described for the eponymous ROI of Fig. 8 and it is placed with respect to the incident laser direction as described in Fig. 13. (a) Concentration profiles for $Ga^{1+}$, $Ga^{2+}$, and N. (b) Concentration profiles for $In^{1+}$ and $In^{2+}$. The results are comparable to those illustrated in Figs. 13 (a, b).

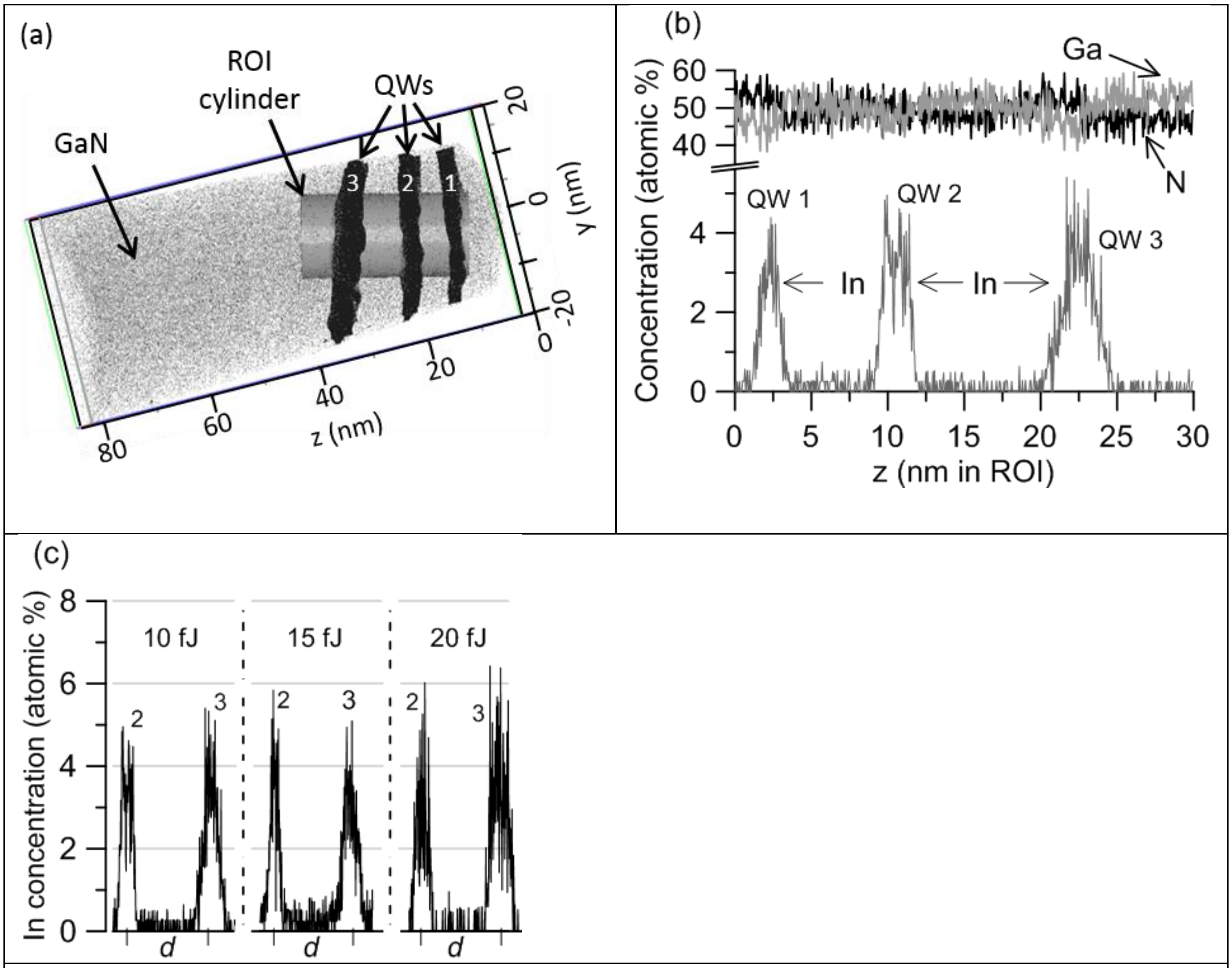

Fig. 15. (a) TLROI of recon 30 for *m*-plane sample with PE = 10 fJ. The three QWs are highlighted with isoconcentration surfaces for indium set to 1.0 at. %. The QWs appear tilted with respect to the reconstruction z-axis because of FIB processing issues. An ROI cylinder (15 nm diameter) is placed centrally with respect to the x-axis and tilted to accommodate the tilt of the QW. (b) Axial concentration profiles for Ga, N, and In along z-axis (long axis) of cylindrical ROI. (c) Comparative axial concentration profiles for In in the bottom two QWs for recons 30, 31, 32, with respective PEs of 10 fJ, 15, fJ and 20 fJ. The concentration profiles are computed within 15-nm-diameter ROI cylinders which are placed and tilted as described in (a). The reconstruction parameters are chosen to yield the QW separation $d \approx 12$ nm for each case. The results indicate that, for this range of PE, the measured PE-dependent variation of In concentration cannot unambiguously be distinguished from possible spatially-varying concentrations of In that occur during growth.

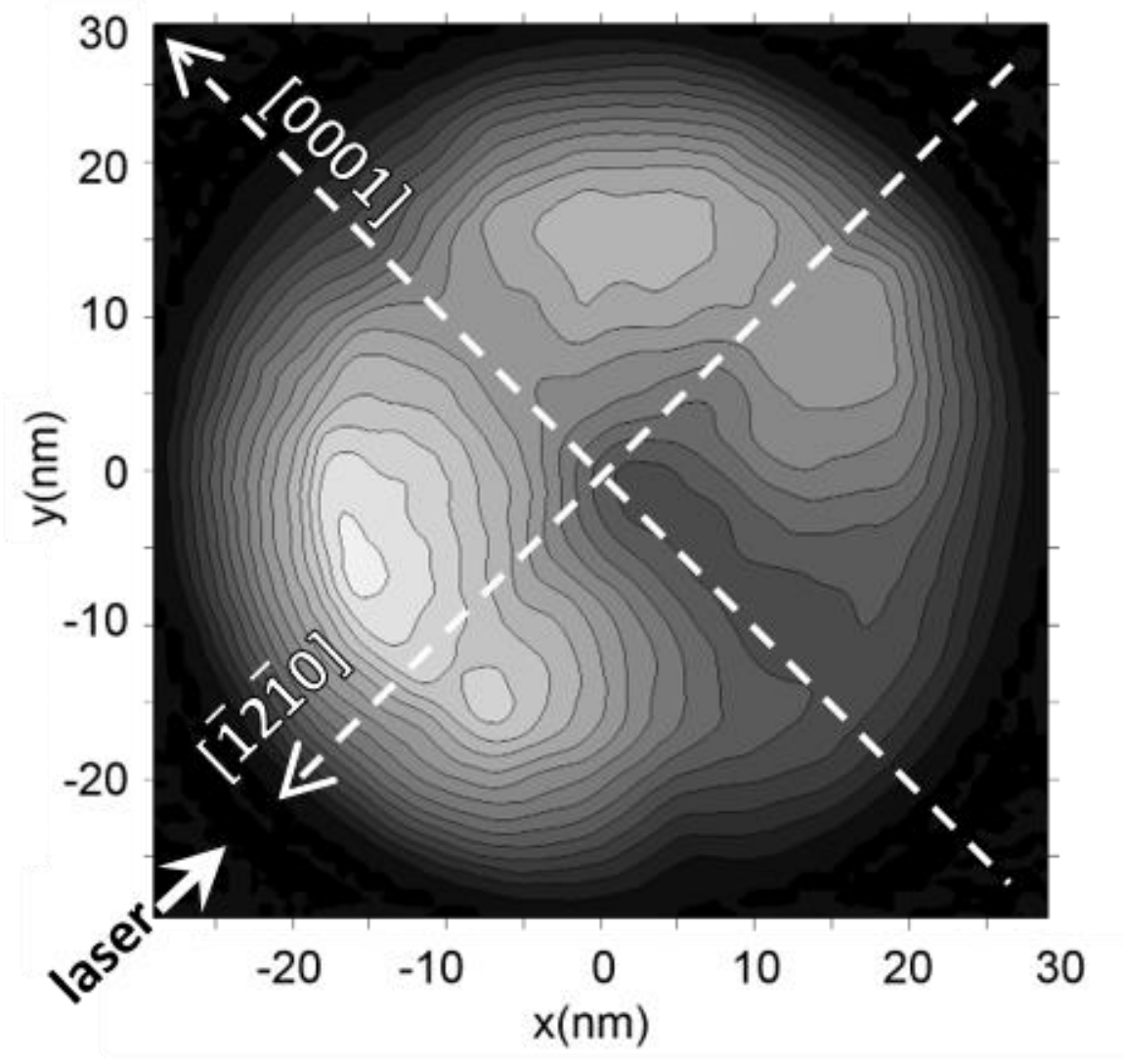

Fig. 16. 2D relative density map of Ga derived from the TLROI of recon 32 with PE = 20 fJ. The indicated axes are established as described in the text.

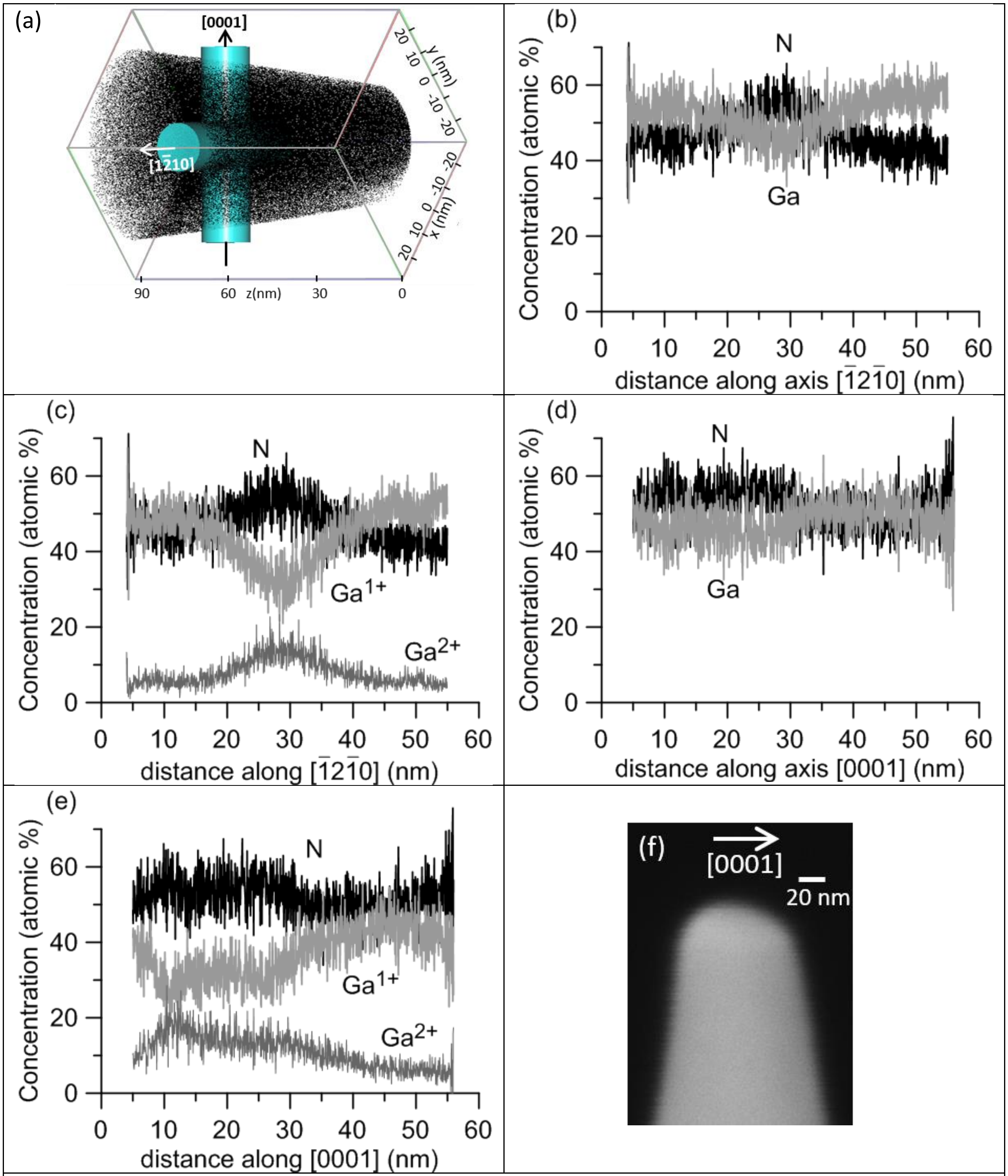

Fig. 17. (a) TLROI of reconstruction 32 showing two ROI cylinders, each 15 nm in diameter, placed along the $[0001]$ and $[\bar{1}2\bar{1}0]$ directions, respectively. The ROIs encompass only GaN and are located roughly 50 nm below the QWs. (b) Concentration profiles for Ga and N within the ROI along $[\bar{1}2\bar{1}0]$. (c) Same as (b) but the Ga concentration is decomposed to show contributions of $Ga^{1+}$ and $Ga^{2+}$. (d) Same as (b) but along $[0001]$. (e) Same as (c) but along $[0001]$. (f) Tip status after ATP has removed all of the QWs. The tip evolves in the GaN with its geometric apex shifting toward -*c*.

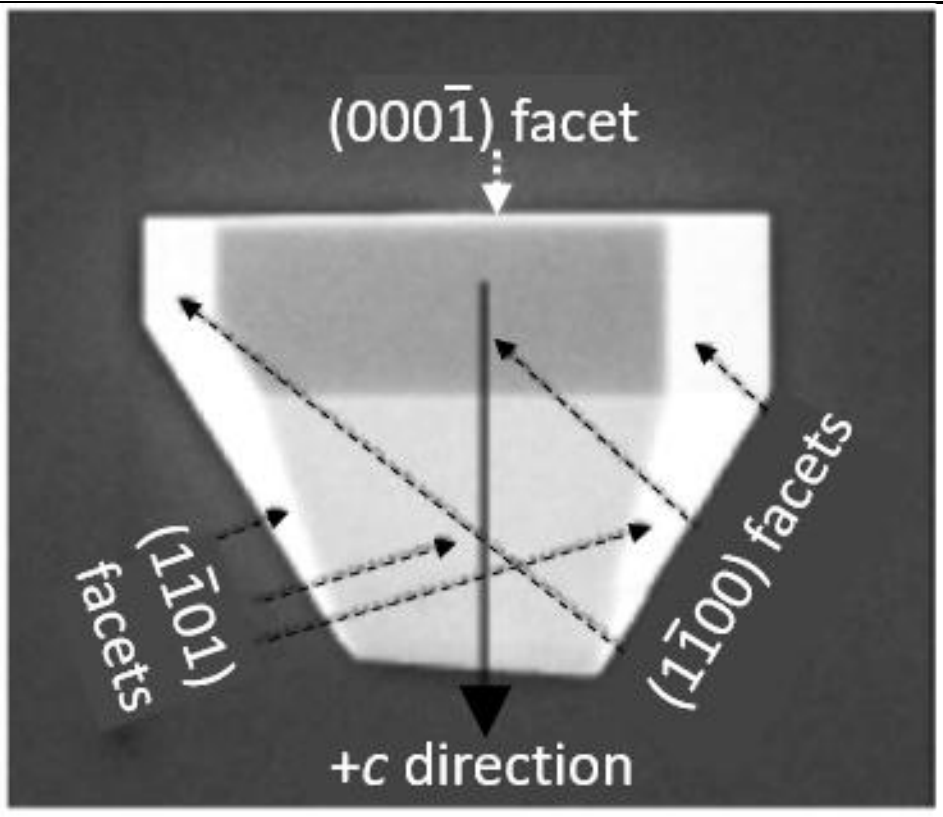

Fig. 18. Growth habit of an *m*-plane, GaN structure from Ref. 38. (Adapted from J. Appl. Phys. 106, 083115 (2009); https://doi.org/10.1063/1.3253575, with the permission of AIP Publishing.) The morphology of the structure, and the etching results of Ref. 39, lend credence to the hypothesis that the transverse shape of the APT tip should evolve asymmetrically with respect to the *c*-axis as illustrated in Fig. 17 (f) and Fig. 19 (a). The image also suggests that the transverse shape of the tip should evolve symmetrically with respect to the *a*-axis as shown in Fig. 19 (b).

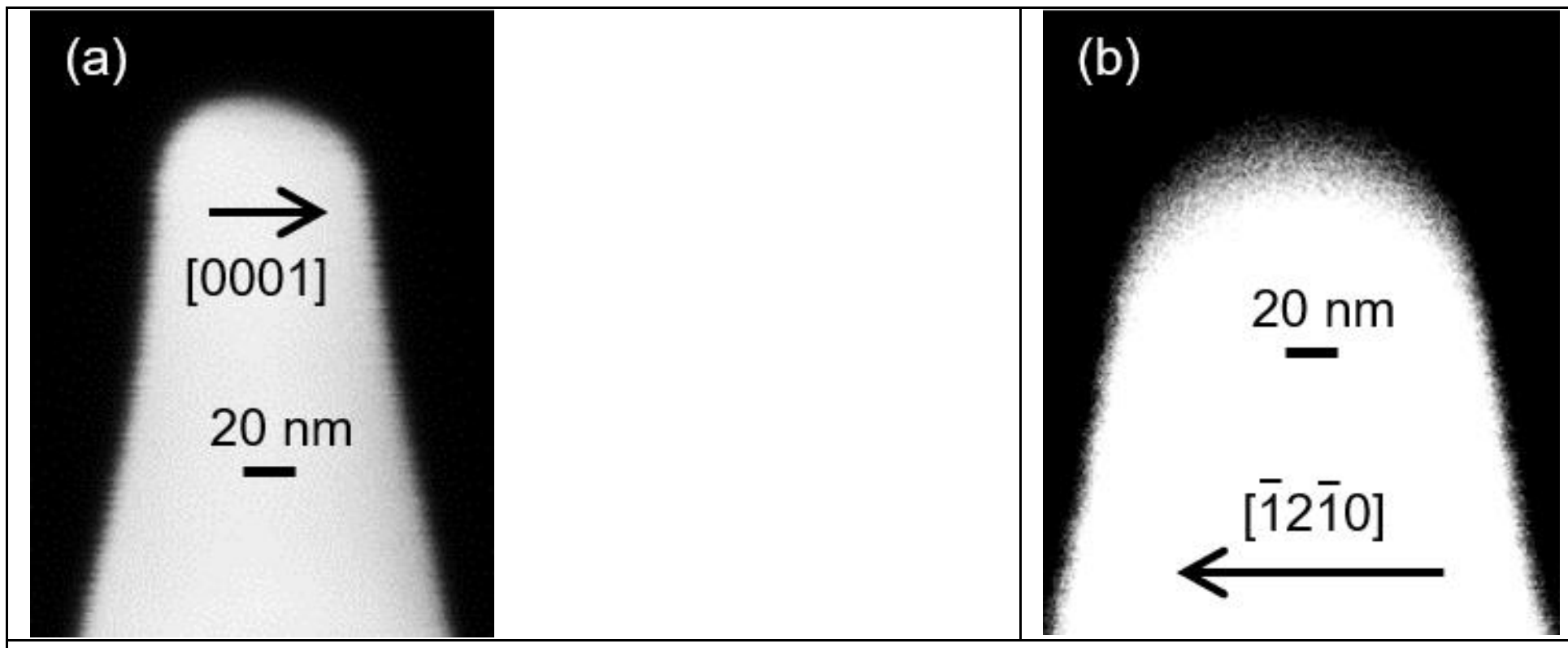

Fig. 19. Orthogonal views of *m*-plane GaN tip from recon 33 after APT analysis. (a) *c*-axis in plane of image showing that the tip evolves such that apex is displaced in the -c direction, which is consistent with Fig. 17 (f). (b) *a*-axis in plane of image showing that the tip evolves symmetrically. The corresponding axial concentration profiles for these cases are quite similar to those shown in Fig. 17 (b—e) and are therefore omitted for brevity.

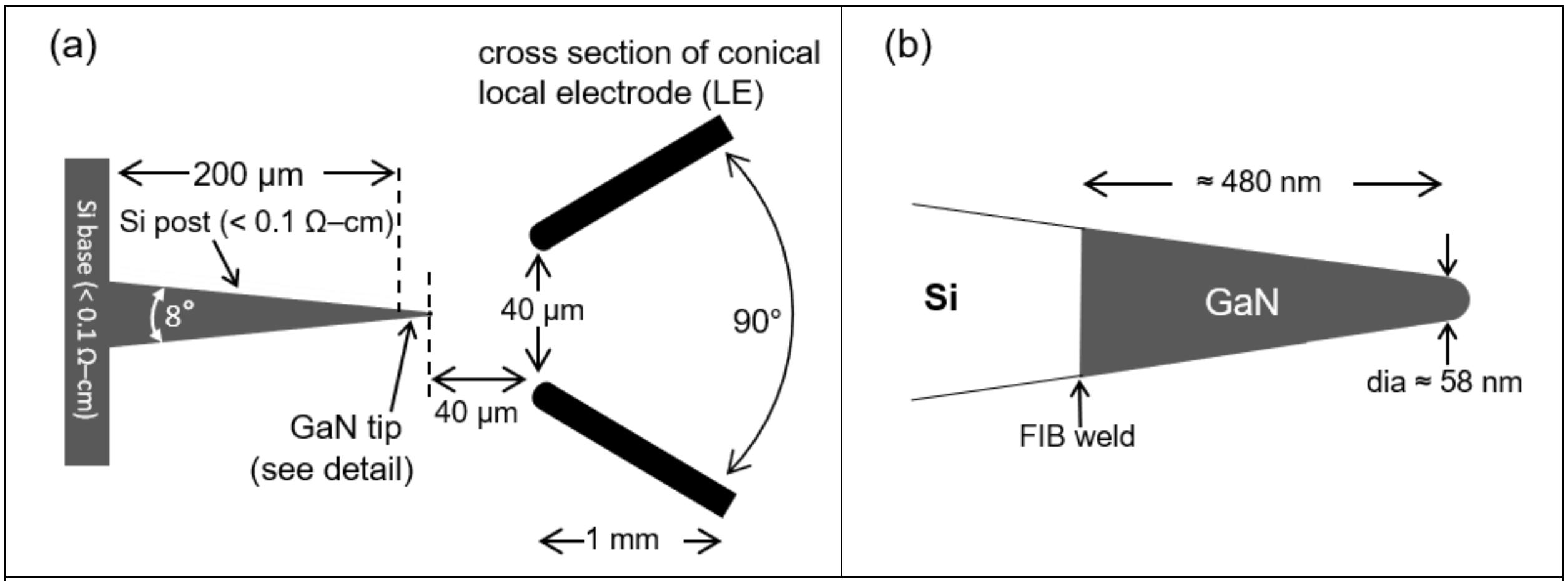

Fig. 20. (a) Schematic showing tip mount and LE for electrostatic simulation of specimen used in recon 32. Si post given the same taper as the GaN tip. The conical LE geometry adapted from Ref. 42. Thickness of LE cone wall = 2 μm; radius of curvature of the wall edge facing tip = 1 μm. Tip—LE separation and post height are representative. (b) Detail showing schematic of tip used in recon 32 at the end of APT data acquisition. The residual GaN length and resulting tip diameter are indicated. For purposes of electrostatic simulations, the FIB weld is assumed to be an Ohmic contact between Si and GaN.

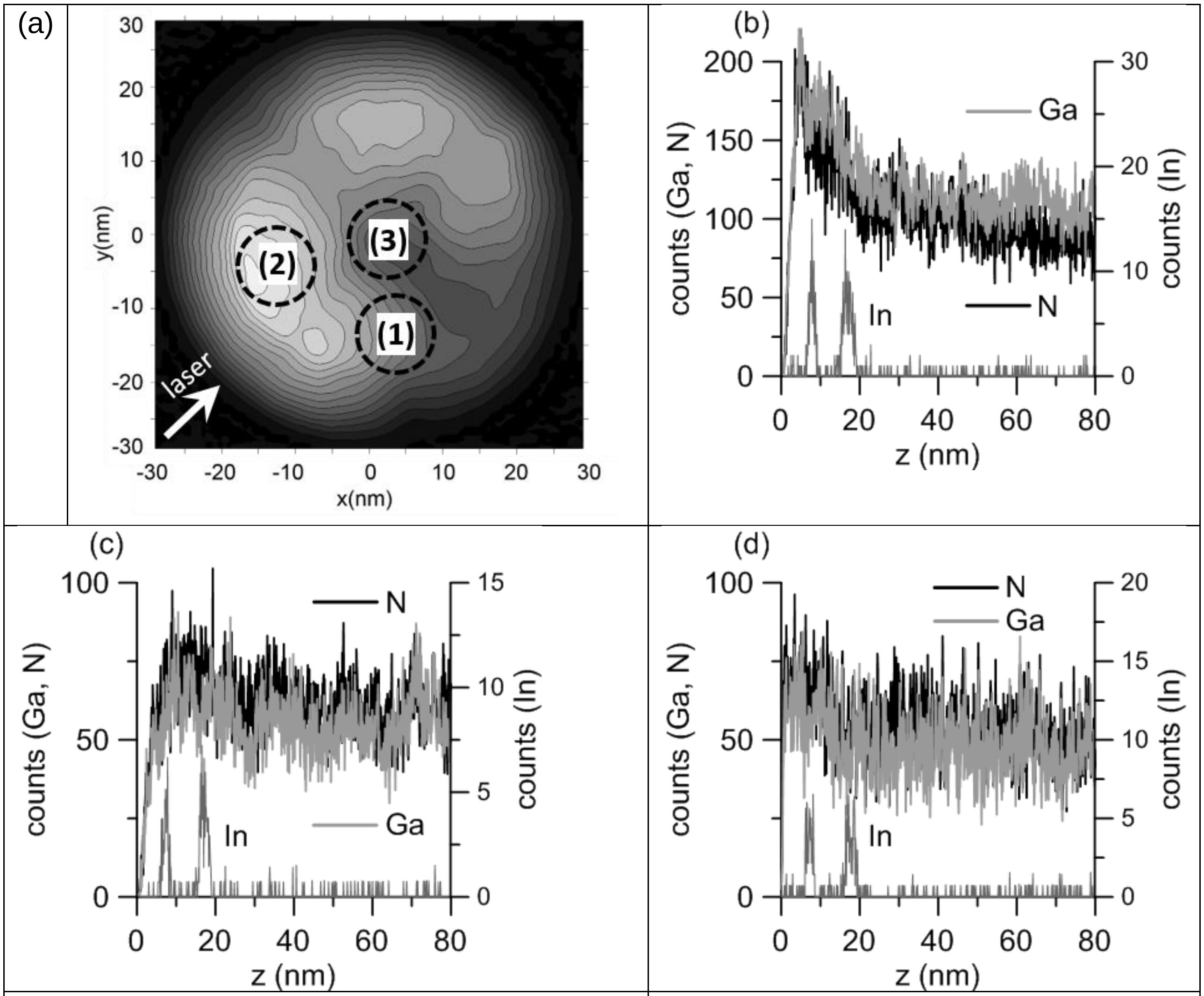

Fig. 21. Spatial dependence of detected counts in *m*-plane sample of recon 32 with PE = 20 fJ. (a) The 2D map of Fig. 16 is used to place ROI cylinders labeled 1—3, each of 10 nm diameter, and oriented parallel to the z axis. The placement and diameter of the ROIs is chosen to approximately conform with similar work described in Ref. 18. (b) Axial dependence of detected counts of Ga, N, and In for ROI 2, (c) ROI 1, and (d) ROI 3. Note the changes in the vertical scales for the three cases. The intersection of the curved tip of the TLROI with ROI cylinders 1 and 2 results in the abrupt rise in Ga and N counts shown in (b) and (c). The artifact is not seen in (d) since ROI 3 is placed at the tip apex. For all cases in (b—d), the bin width per depth increment is 0.1 nm.

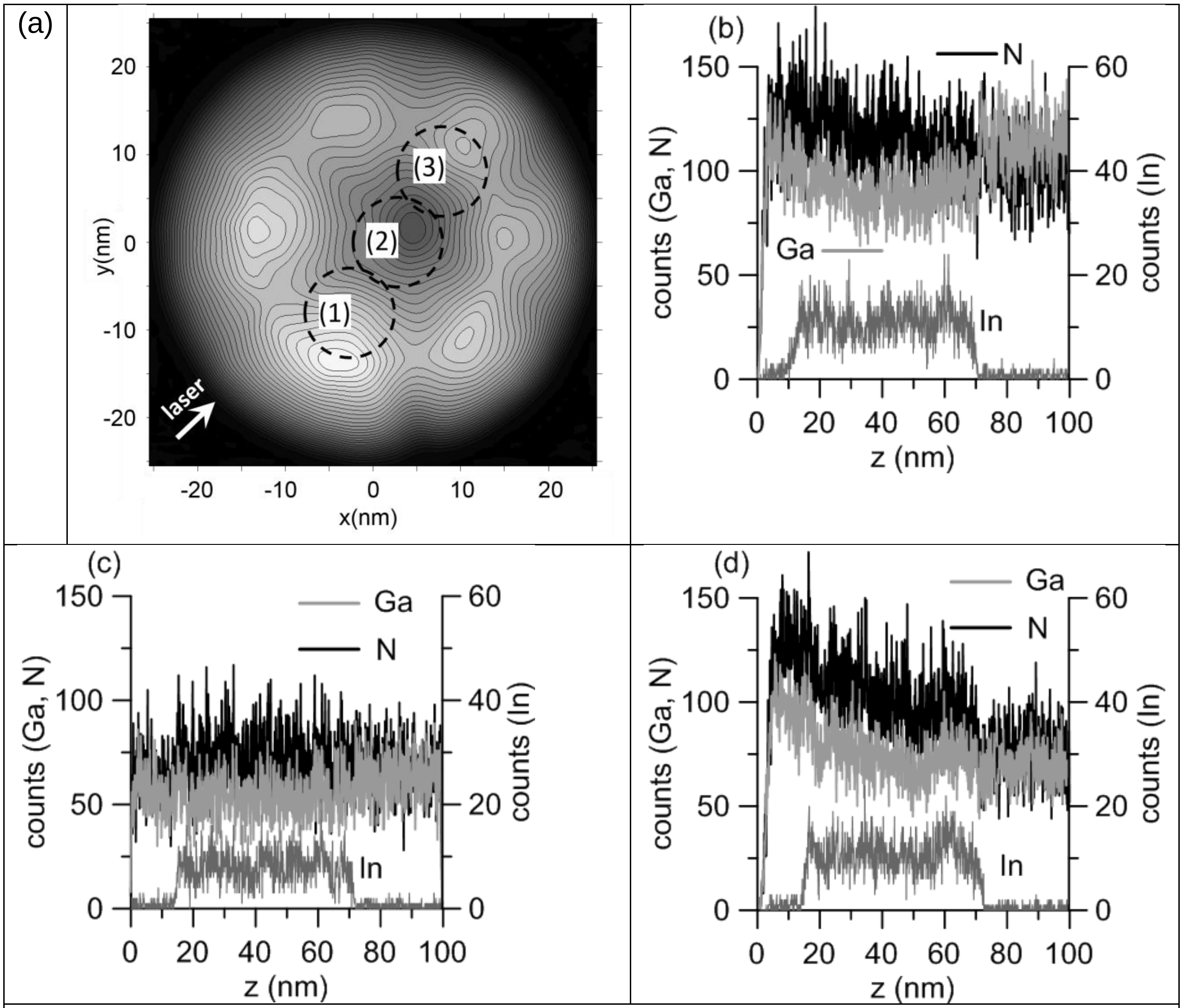

Fig. 22. Spatial dependence of detected counts in *c*-plane sample of recon 3 with PE = 10 fJ. (a) 2D relative density map for Ga (similar to that of Fig. 7) is used to place ROI cylinders labeled 1—3, each of 10 nm diameter, and oriented parallel to the z axis. (b) Axial dependence of detected counts of Ga, N, and In for ROI 1, (c) ROI 2, and (d) ROI 3. Note the changes in the vertical scales for the three cases. The intersection of the curved tip of the TLROI with ROI cylinders 1 and 3 results in the abrupt rise in Ga and N counts shown in (b) and (d). The artifact is not seen in (c) since ROI 2 is placed at the tip apex. For all cases in (b—d), the bin width per depth increment is 0.1 nm.